\documentclass[namedreferences]{solarphysics}
\usepackage[optionalrh]{spr-sola-addons} % For Solar Physics 
\usepackage{graphicx}        % For eps figures, newer & more powerfull
\usepackage{color}           % For color text: \color command

\usepackage[hyphens]{url}             % For breaking URLs easily trough lines
\usepackage[pdfborder={0 0 0 },urlcolor=blue,breaklinks]{hyperref} \ifx \doiurl \undefined \def \doiurl#1{\href{http://dx.doi.org/#1}{\url{#1}}}\fi \ifx \adsurl \undefined \def \adsurl#1{\href{http://adsabs.harvard.edu/abs/#1}{\url{#1}}}\fi

\newcommand{\etal}{{\it et al.}}

\newcommand{\aap}{    {\it Astron. Astrophys.}}

\newcommand{\apj}{    {\it Astrophys. J.}}
\newcommand{\apjl}{   {\it Astrophys. J. Lett.}}

\newcommand{\mnras}{  {\it Mon. Not. Roy. Astron. Soc.}}
\newcommand{\nat}{    {\it Nature}}

\newcommand{\solphys}{{\it Solar Phys.}}
 
\newcommand{\ssr}{    {\it Space Sci. Rev.}}

\newcommand{\be}{\begin{equation}}
\newcommand{\ee}{\end{equation}}
\newcommand{\beq}{\begin{eqnarray}}
\newcommand{\eeq}{\end{eqnarray}}

\newcommand{\vect}[1]{{\bf #1}}

\newcommand{\diff}{{\mathrm d}}

\renewcommand{\inlinecite}[1]{\citeauthor{#1} (\citeyear{#1})}

\begin{document}
\begin{article}
\begin{opening}
\title{Properties of the 15 February 2011 Flare Seismic Sources.}            
%\subtitle{Seismic properties}

\author{S.~\surname{Zharkov}$^{1}$\sep
	L.M.~\surname{Green}$^{1}$\sep 
	S.A.~\surname{Matthews}$^{1}$\sep
	V.V.~\surname{Zharkova}$^{2}$
	}
\institute{$^{1}$ UCL Mullard Space Science Laboratory,
          University College London, Holmbury St. Mary, Dorking, RH5 6NT  UK
          email:\url{sz2[at]mssl.ucl.ac.uk} \\                              
	$^{2}${Department of Mathematics,} University of Bradford, Bradford, {BD7 1DP} UK
	}

%\date: rather not

%\translation{De Kluwer LaTeX stylefile; aanwijzingen voor auteurs}

\runningtitle{15 February 2011 Flare: Seismic Properties}
\runningauthor{S.~Zharkov \etal}

\begin{abstract} 
The first near-side X-class flare of the Solar Cycle 24 occurred in February 2011 and produced a very strong seismic response in the photosphere. One sunquake was reported by 
%\inlinecite{Kosovichev2011}, 
\citeauthor{Kosovichev2011} (\apjl\ {\bf 734},  {L15}, \citeyear{Kosovichev2011}),
followed by the discovery of a second sunquake by 
\citeauthor{ZGMZ2011} (\apjl\ {\bf 741}, {L35}, \citeyear{ZGMZ2011}).
%\inlinecite{ZGMZ2011}.
The flare had a two-ribbon structure and was associated with a flux rope eruption and a halo coronal mass ejection (CME) as reported in the CACTus catalogue.
Following the discovery of the second sunquake and the spatial association of both sources with the locations of the feet of the erupting flux rope 
(\citeauthor{ZGMZ2011}\apjl\ {\bf 741}, {L35}, \citeyear{ZGMZ2011}).
%\cite{ZGMZ2011}, 
we present here a more detailed analysis of the observed photospheric changes in and around the seismic sources. 
These sunquakes are quite unusual, taking place early in the impulsive stage of the flare, with the seismic sources showing little hard X-ray (HXR) emission{, and strongest X-ray emission sources  located in the flare ribbons.}
%{\em
%The strongest X-ray emission sources are located in the flare loops and so we also consider the photospheric changes there in comparison to the locations of the two seismic sources. }
We present a directional time--distance diagram computed for the second source, which clearly shows a ridge corresponding to the travelling acoustic wave packet and find that the quake at the second source happened about 45 seconds to one minute earlier than the first source. Using acoustic holography we report different frequency responses of the two sources. We find strong downflows at both seismic locations and a supersonic horizontal motion at the second site of acoustic wave excitation.
\end{abstract}


\abbreviations{\abbrev{SDO}{Solar Dynamics Observatory};\\
   \abbrev{HMI}{Heliographic Michelson Imager};
   \abbrev{AIA}{Atmospheric Imaging Assembly}}

%	\nomenclature{\nomen{KAP}{Kluwer Academic Publishers};
%	   \nomen{compuscript}{Electronically submitted article}}

%	\classification{JEL codes}{D24, L60, 047}
\end{opening}

\section{Introduction}
\label{sec:intro}

Sunquakes are observed as photospheric ripples, which accelerate radially outward from a source region. The theoretical prediction that sunquakes should be produced by the energy released during major solar flares \cite{wolff72} was supported by their discovery on the Sun by \inlinecite{kz1998}. The acoustic nature of quakes has been well established since their discovery. However, the exact physical mechanism behind their excitation is still debated with several theories currently under consideration. Observations of sunquakes are relatively rare, possibly due to the difficulties of detecting the photospheric ripples, and helioseismic methods such as time--distance diagram analysis and acoustic holography are employed to look for evidence of acoustic emission. 
With only a small number of such events verified so far (see \opencite{Besliu2005}; \opencite{Donea2006list} for some examples of known quakes from the last solar cycle) the new solar cycle and the virtually continuous high-resolution data  of SDO/HMI \cite{ScherrerHMI2011,Schou2011hmi} mean that sunquake detections should increase in the coming years.

The 15 February 2011, X2.2 class flare occurred in NOAA active region 11158 and was the first in the much delayed rising activity phase of the new Solar Cycle 24. The active region started emerging in the eastern hemisphere on 10 February 2011, with two bipoles emerging side by side creating a complex multipolar region. As the active region evolved through both emergence and cancellation events, the coronal loops became increasingly sheared resulting in a number of C-class and  M-class flares occurring from 13 February onward, culminating in the X-class event with GOES flux peaking around 01:55\,UT on 15 February. The X-class flare was a long-duration flare with an impulsive phase, as observed in GOES 1.0 to 0.8 \AA \, soft X-ray data, lasting from 01:46 to 01:56\,UT, and integrated HXR emission observed by {RHESSI} up to $\approx 100 {\rm \ keV}$, peaking just before 01:55\,UT. The strongest HXR emission was produced in the $6\,-\,25{\rm \ keV}$ energy band.

\begin{figure*}
\centerline{
\begin{tabular}{c @ {\hspace{1pc}} c} % @ {\hspace{1pc}} c}
\color{black}{(a)} \includegraphics[width=.5\textwidth]{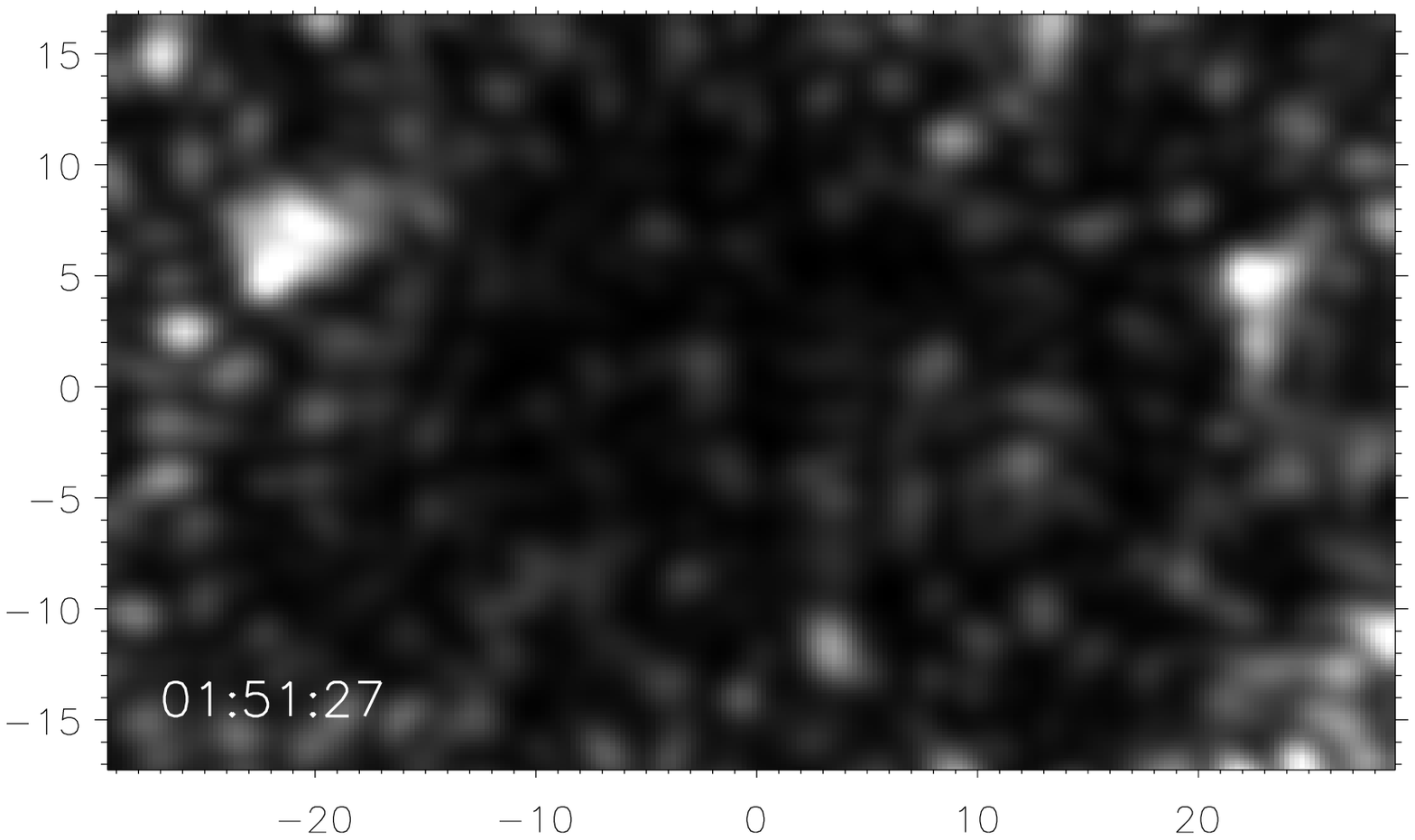}  &
\color{black}{(b)}\includegraphics[width=.5\textwidth]{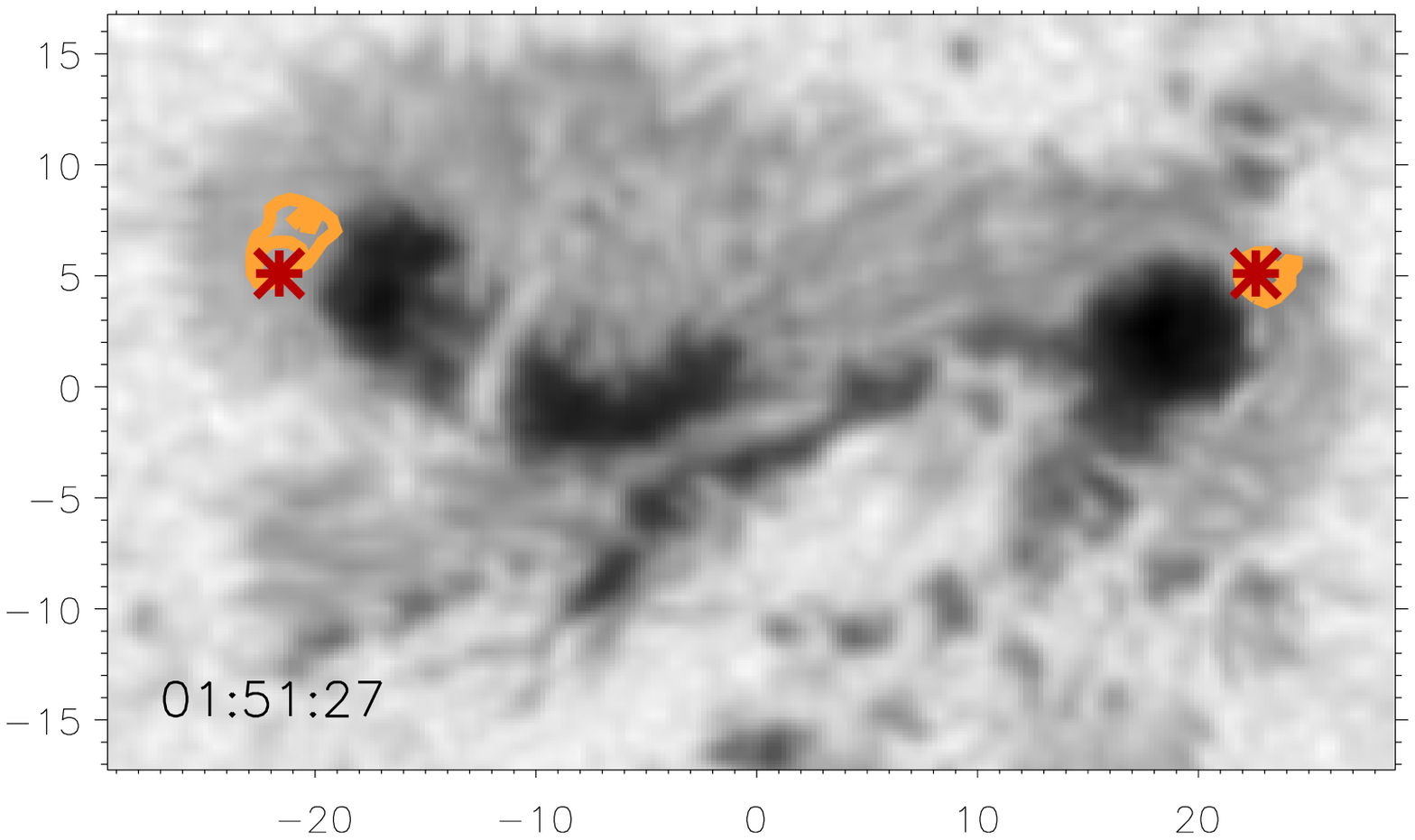} \\  
\color{black}{(c)} \includegraphics[width=.5\textwidth]{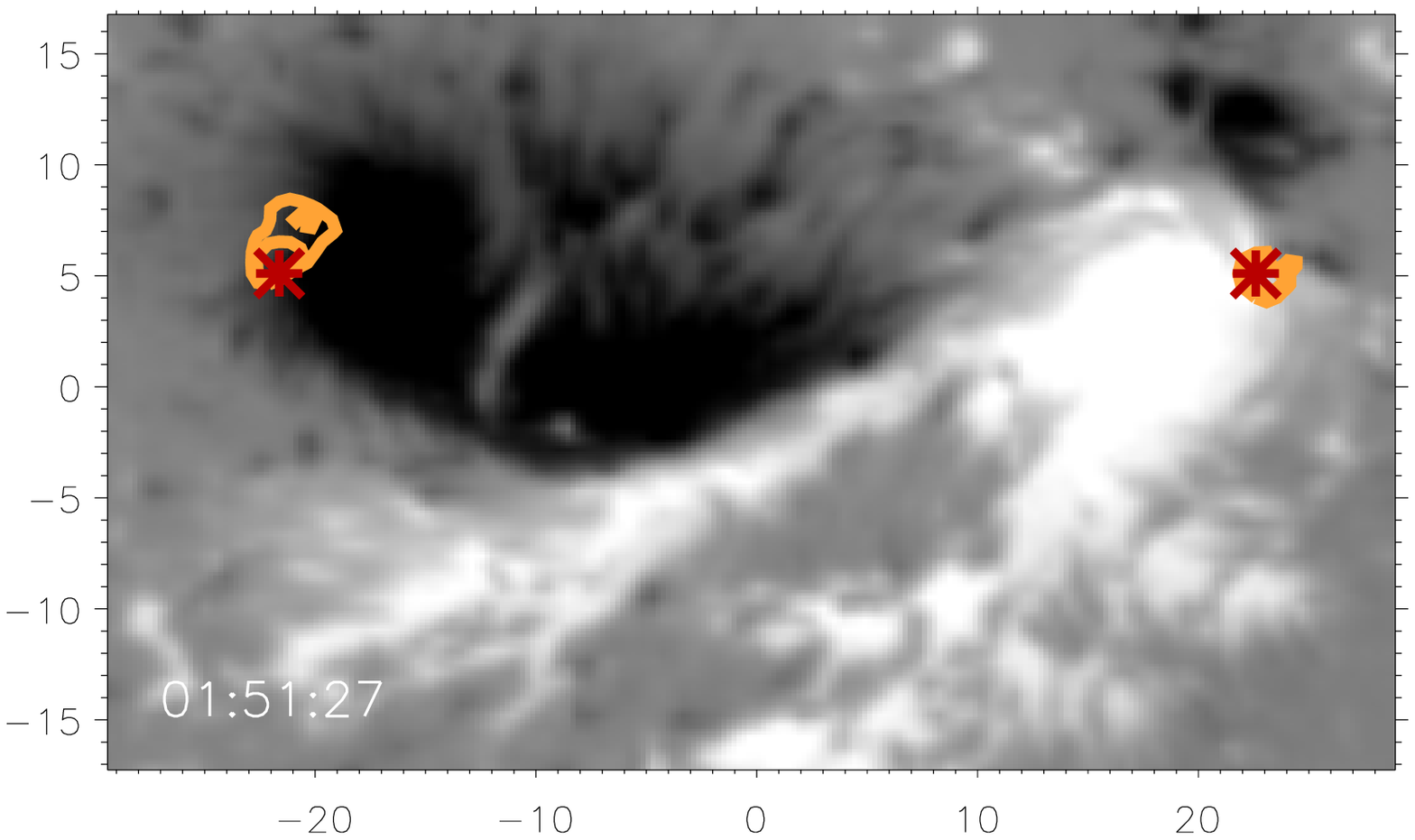} & 
  %    {\Large \bf   
  %    \hspace{0.0 \textwidth}  \color{white}{(a)}
   %   \hspace{0.415\textwidth}  \color{white}{(b)}
 %        \hfill}
\color{black}{(d)}\includegraphics[width=.5\textwidth]{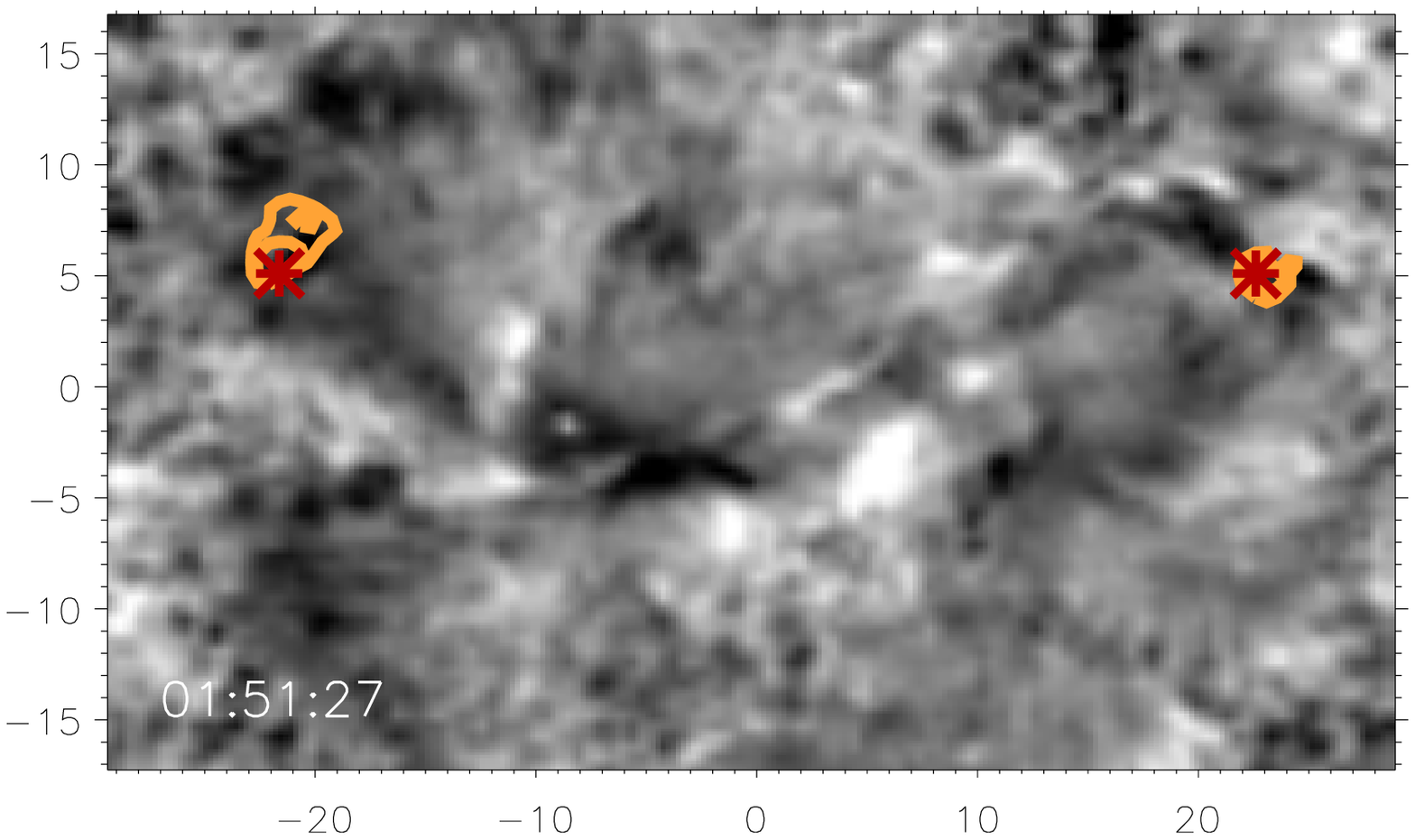}\\ 
\color{black}{(e)}\includegraphics[width=.5\textwidth]{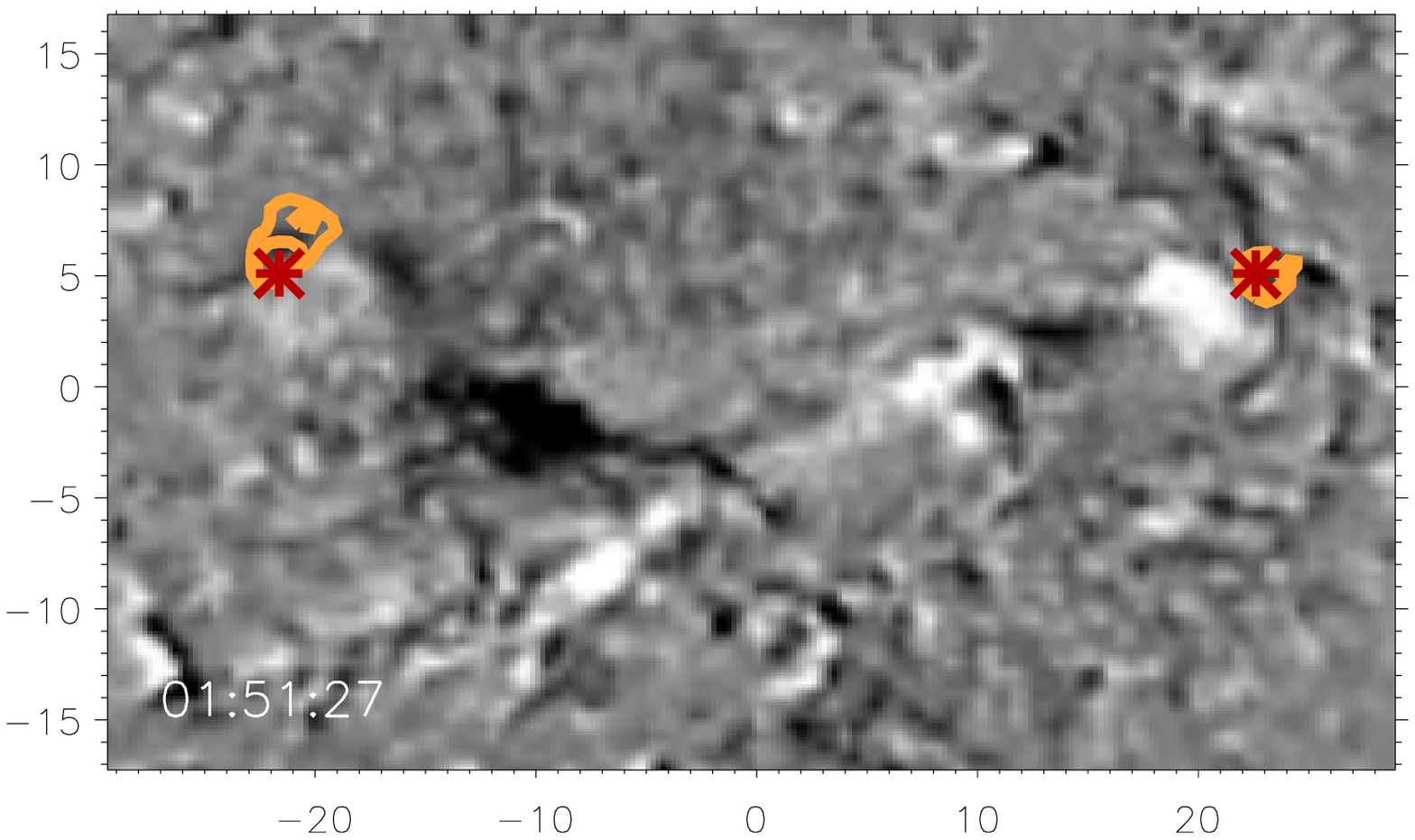} & 
\color{black}{(f)} \includegraphics[width=.5\textwidth,bb=50 65 410 280, clip=]{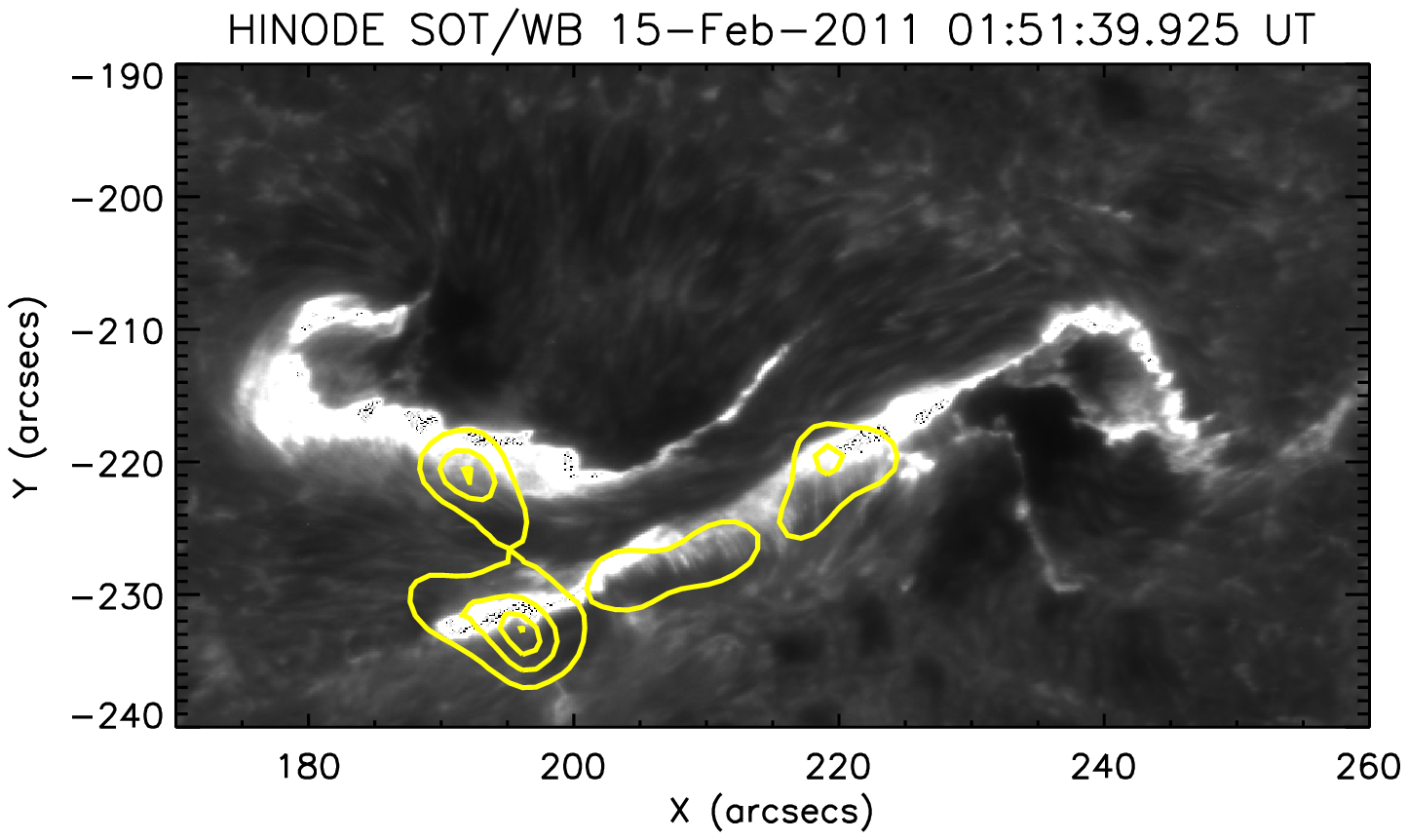} %{sphys_20110215_id_boxes} \\
\end{tabular}
}
\caption{Sunquake source locations determined from egression power computed at 6 mHz and time--distance overplotted on egression power snapshot (a), continuum intensity (b), magnetogram (c) and velocity (d) images. Panel (e) shows the flare-induced changes in magnetic field computed as the difference between magnetic data in panel (c) and the 20-minute average of HMI line-of-sight magnetorgrams before the flare onset.
%The boxes used for mosaic plots are overplotted on intensity difference (f) and magnetic change (e) image at around 01:50 UT. 
The orange contours in (a)\,-\,(e) correspond to 01:51:27 6-mHz egression power contours at 2.5 and 3 times the quiet-Sun egression at this frequency. The red stars in all images mark the locations used for computing time--distance diagrams. Panel (f) is Ca {\sc ii} K taken by {\it Hinode}/SOT at  01:51:39\,UT with HXR contours deduced from RHESSI data in 25\,-\,50 keV range overplotted. The HMI data in panels (a)-(e) are Postel projected so the distances along $x$- and $y$-axes are plotted in megameters. Arcsecond coordinates are given along the axes in panel (f).
}
\label{fig:srcs_boxes}
\end{figure*}

\begin{sloppypar}
In the photosphere, the flare exhibited a classic two-ribbon pattern, which is most clearly seen in the SDO/HMI line-of-sight magnetic field and velocity running-difference images as well as {\em Hinode}/SOT Ca {\sc ii} K observations ({\it e.g.} Figure \ref{fig:srcs_boxes}). Though less pronounced, the ribbons are also present in the SDO/HMI continuum data. Spatially, hard X-ray emission was situated  primarily along the ribbons.  The halo CME associated with this flare was detected by CACTus software and is listed in the LASCO catalogue (\url{http://sidc.oma.be/cactus/catalog/LASCO/2_5_0/qkl/2011/02/}).
\end{sloppypar}

The flare produced a strong seismic response \cite{Kosovichev2011}, with ripples travelling outward from the source clearly seen in HMI velocity difference data (see, for instance, the online movie in the above article). Using acoustic holography  \inlinecite{ZGMZ2011} have shown that a second, apparently weaker, source of acoustic waves is present and have shown that the sunquakes occurred at the foot-points of a flux rope. The presence and location of the flux rope has been deduced based on a straight-forward observational case, making use of the photospheric magnetic flux distribution, sigmoidal structure, chromospheric and coronal changes during the eruption. The observational interpretation of the presence of a flux rope is supported by non-linear magnetic-field modelling from HMI vector magnetogram data \cite{Schrijver2011,XSun2012}. 
%and Xudong Sun - private communication).

In this article we analyse seismic measurements and report the properties of the detected seismic sources and consider associated changes in photosphere. The data and methods are described in Section \ref{sec:observation}, results are presented in Section \ref{sec:results}, followed by discussion and conclusions.

%*******************************************************************************************************************

\begin{figure*}
\centerline{
\includegraphics[width=.5\textwidth]{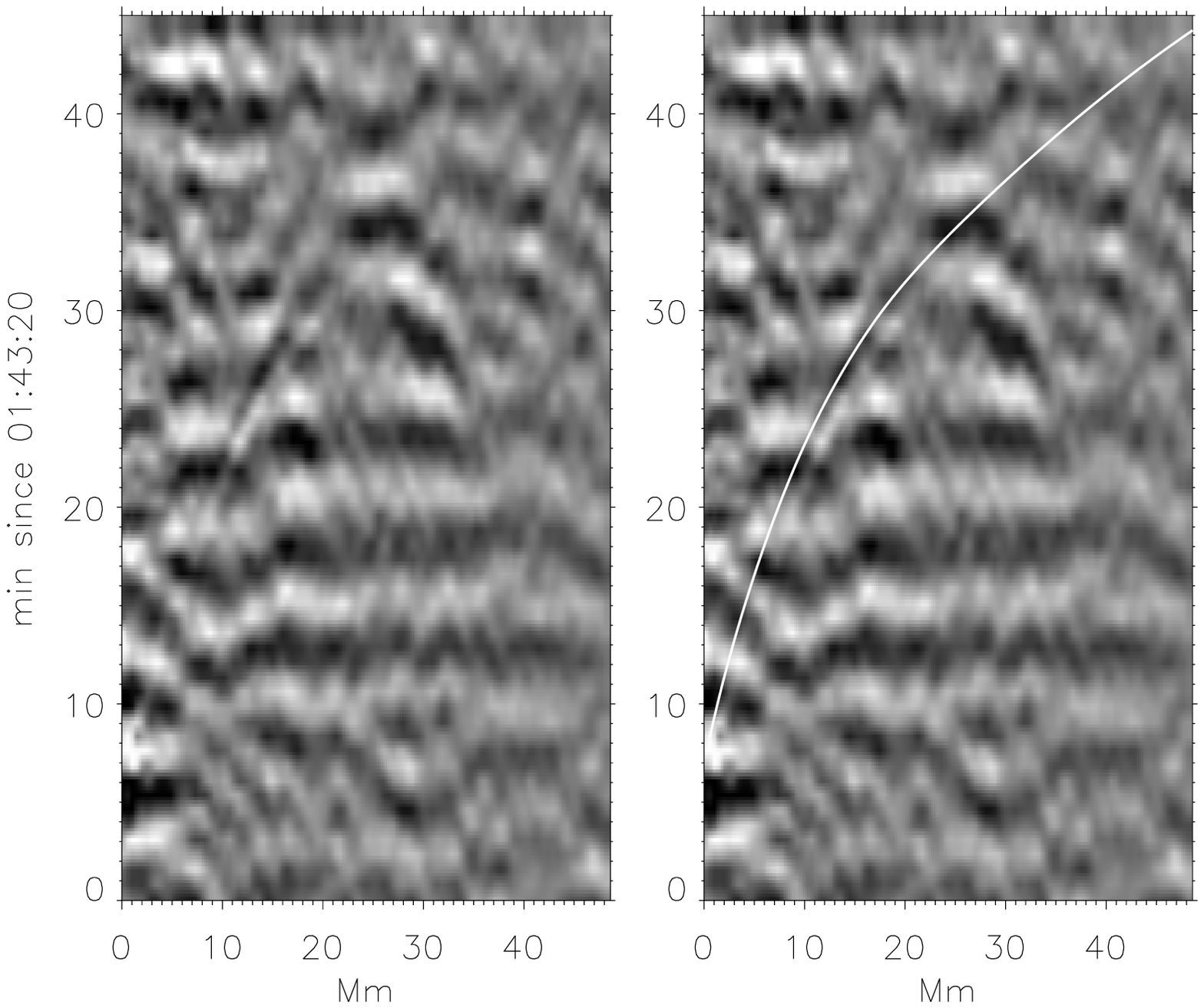}\\
\includegraphics[width=.5\textwidth,bb=15 0 500 385, clip=]{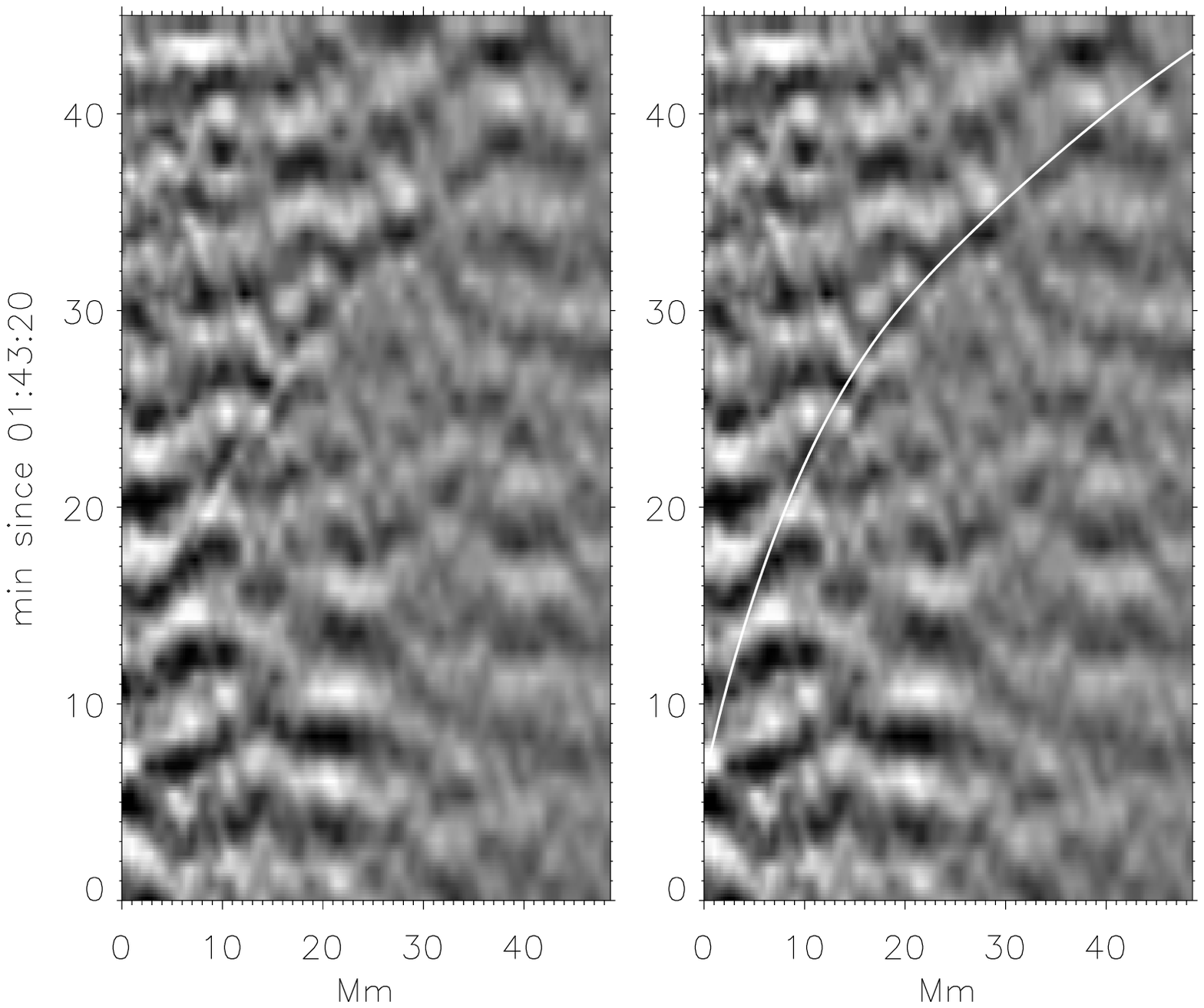}
}
\caption{Time--distance diagrams for both seismic sources. The first two plots from the left correspond to the first (eastern, strong) source, the following are the western source. Theoretical time--distance curves are overplotted in white. Locations of the sources are marked as red stars in Figure \ref{fig:srcs_boxes}.
}
\label{fig:td_s2}
\end{figure*}

%*******************************************************************************************************************

\section{Data and Methods}
\label{sec:observation}

%*******************************************************************************************************************
\begin{figure*}
\centerline{
\begin{tabular}{c @ {\hspace{1pc}} c}
\includegraphics[width=.5\textwidth]{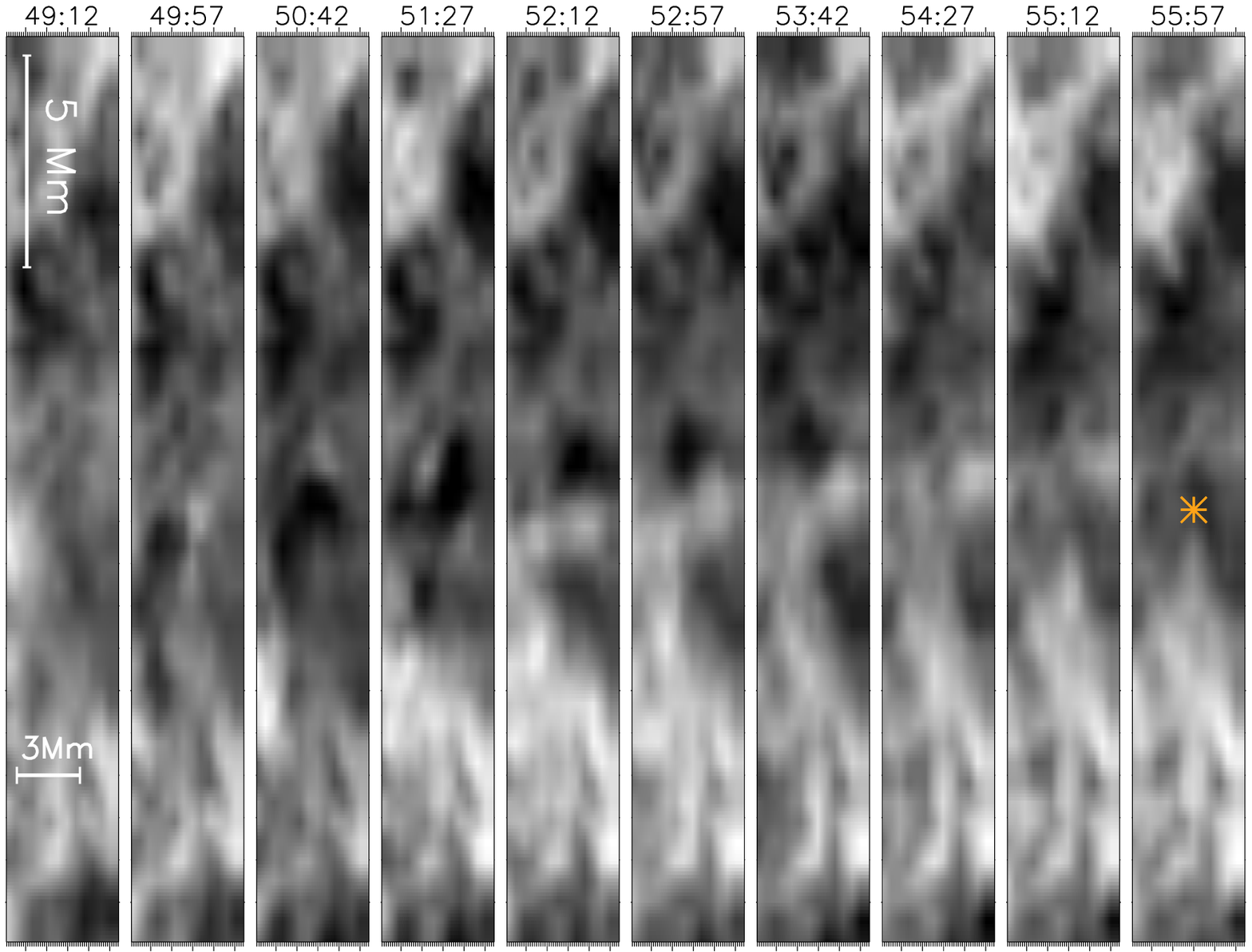} & 
\includegraphics[width=.5\textwidth]{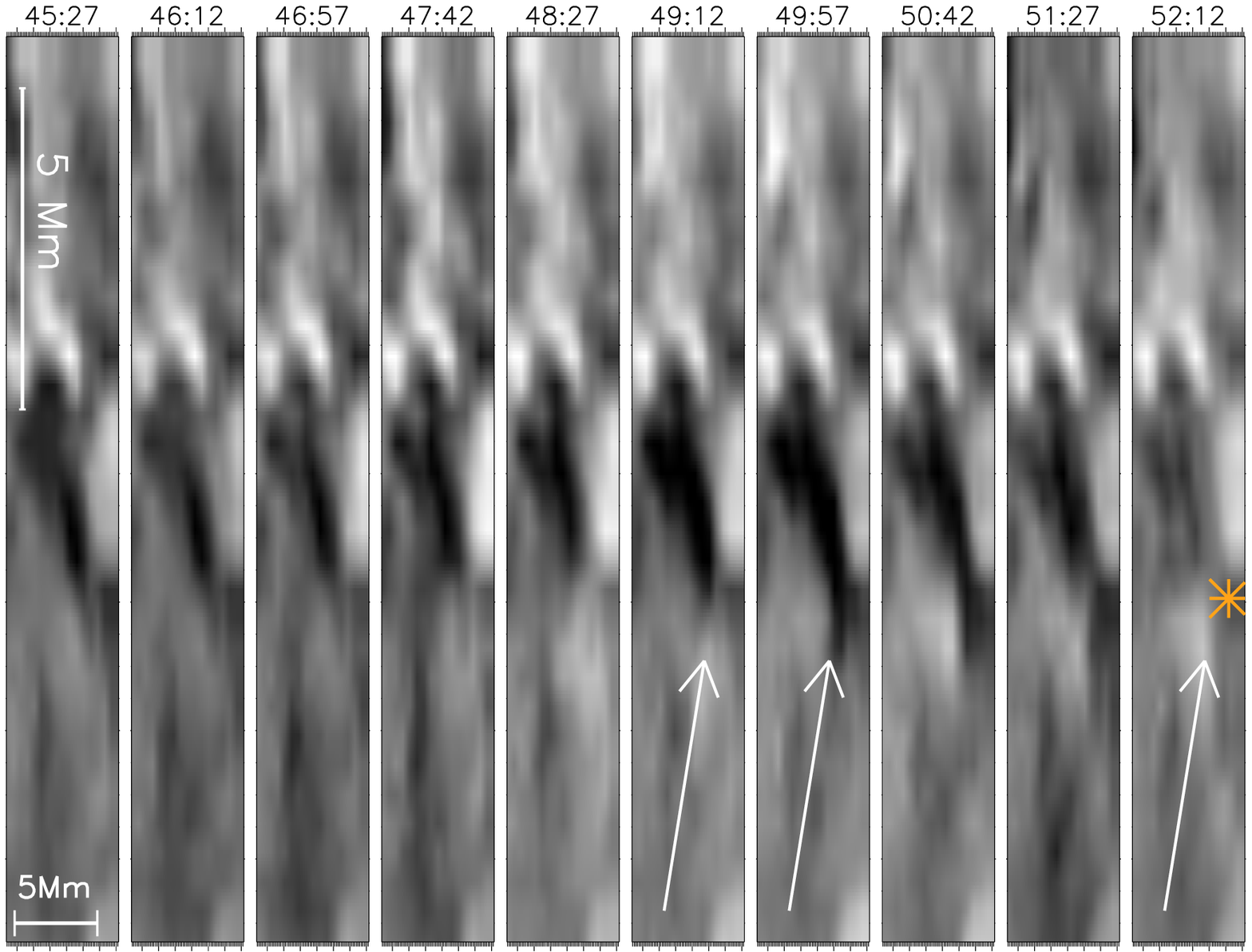} \\
\includegraphics[width=.58\textwidth]{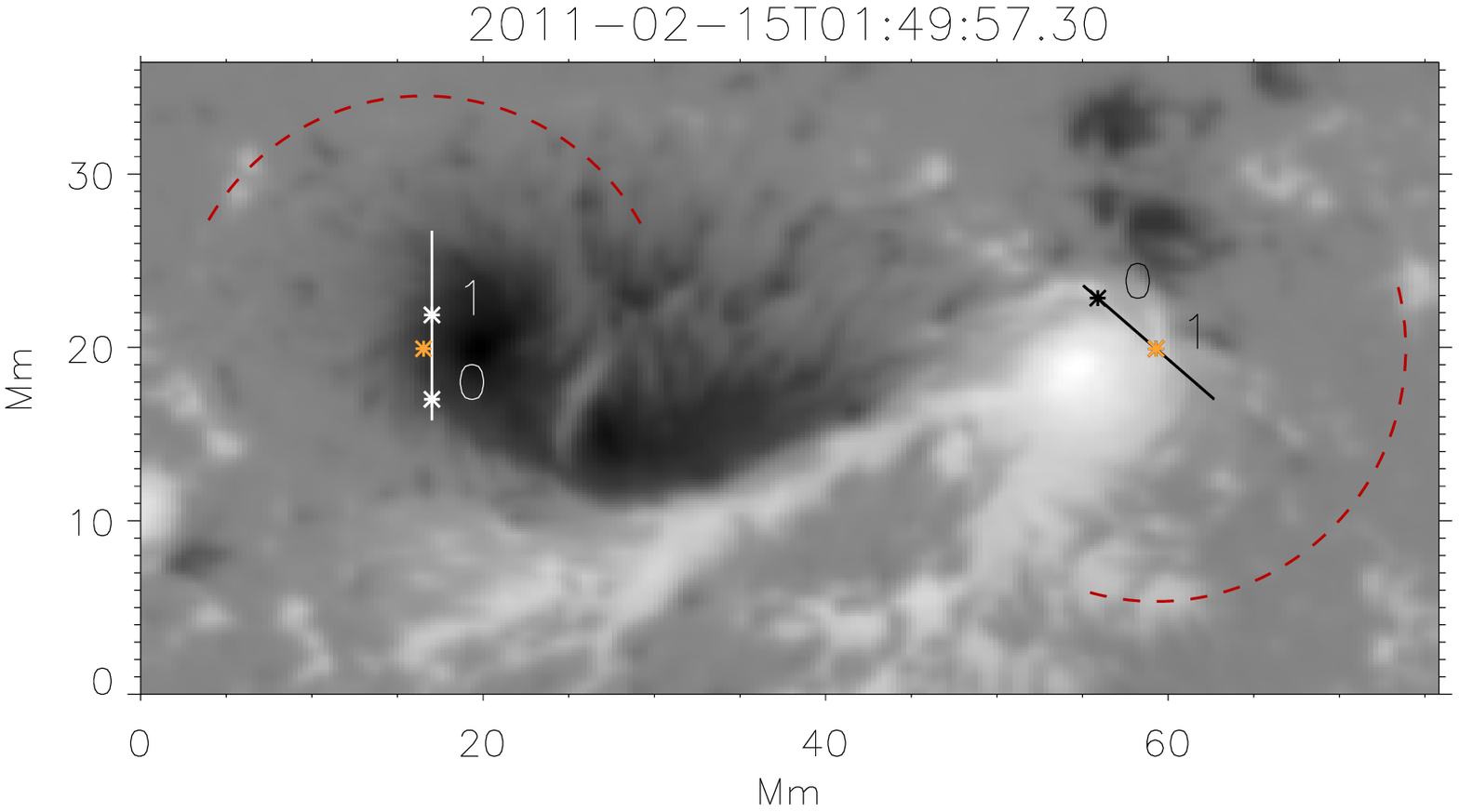} & 
\includegraphics[width=.58\textwidth]{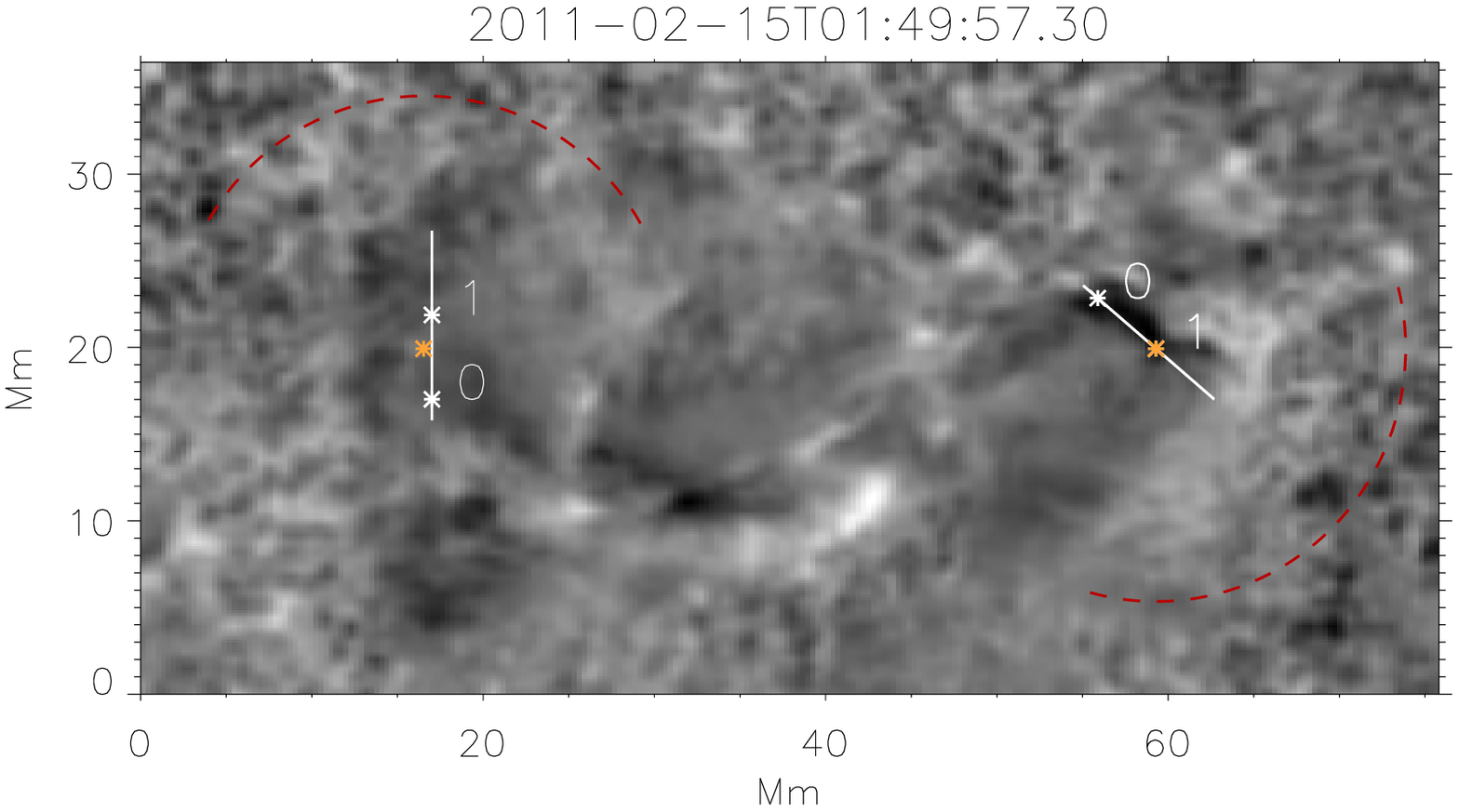}\\ %{20110215_analysis_s1transient_md_context}\\%{sphys_vv_variat_sscale_xxx_dims}\\
%{sphys_vv_variat_sscale_2_dims}\\%{20110215_analysis_s2transient_doppler_line}\\
\includegraphics[width=.58\textwidth]{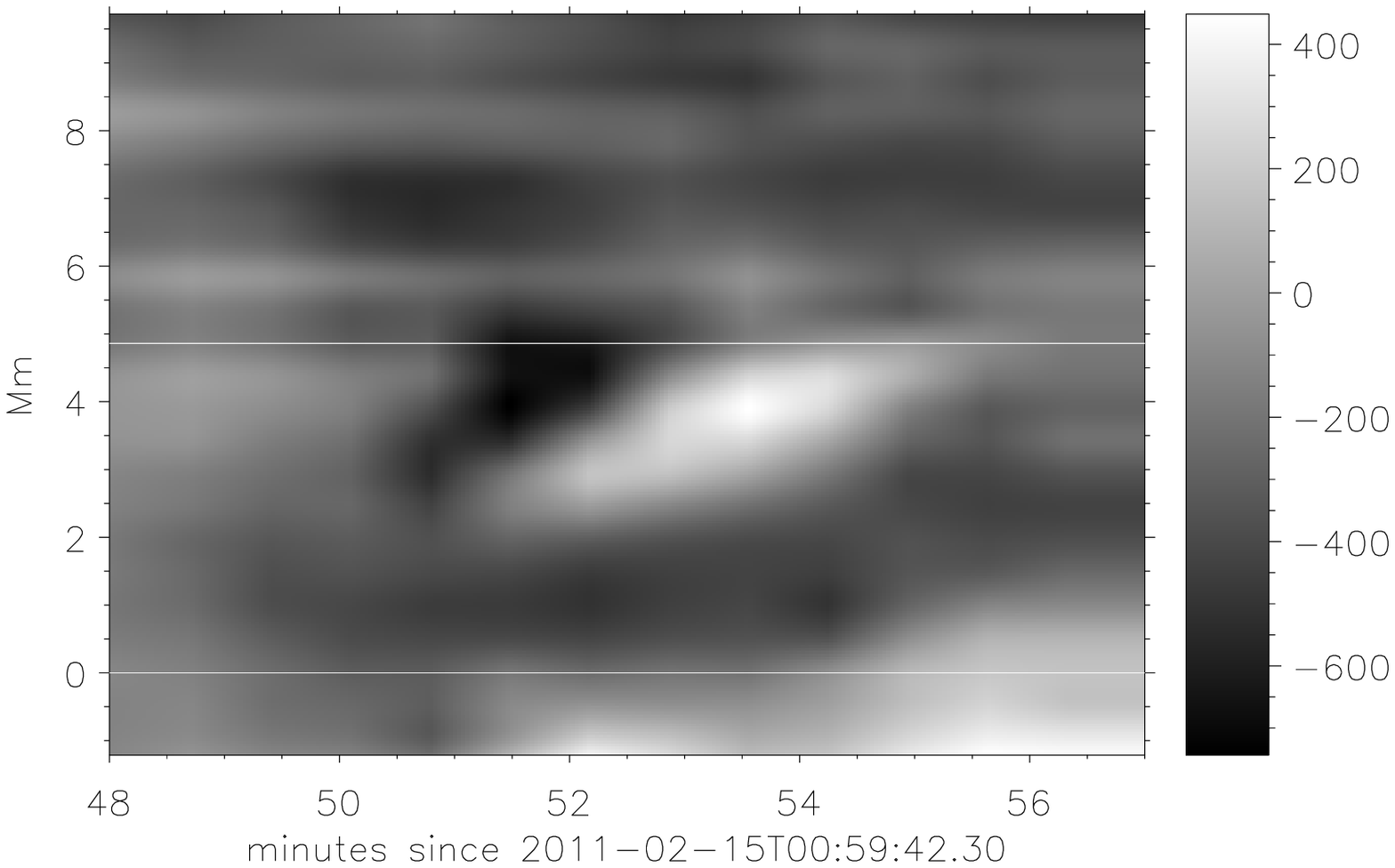} & 
\includegraphics[width=.58\textwidth]{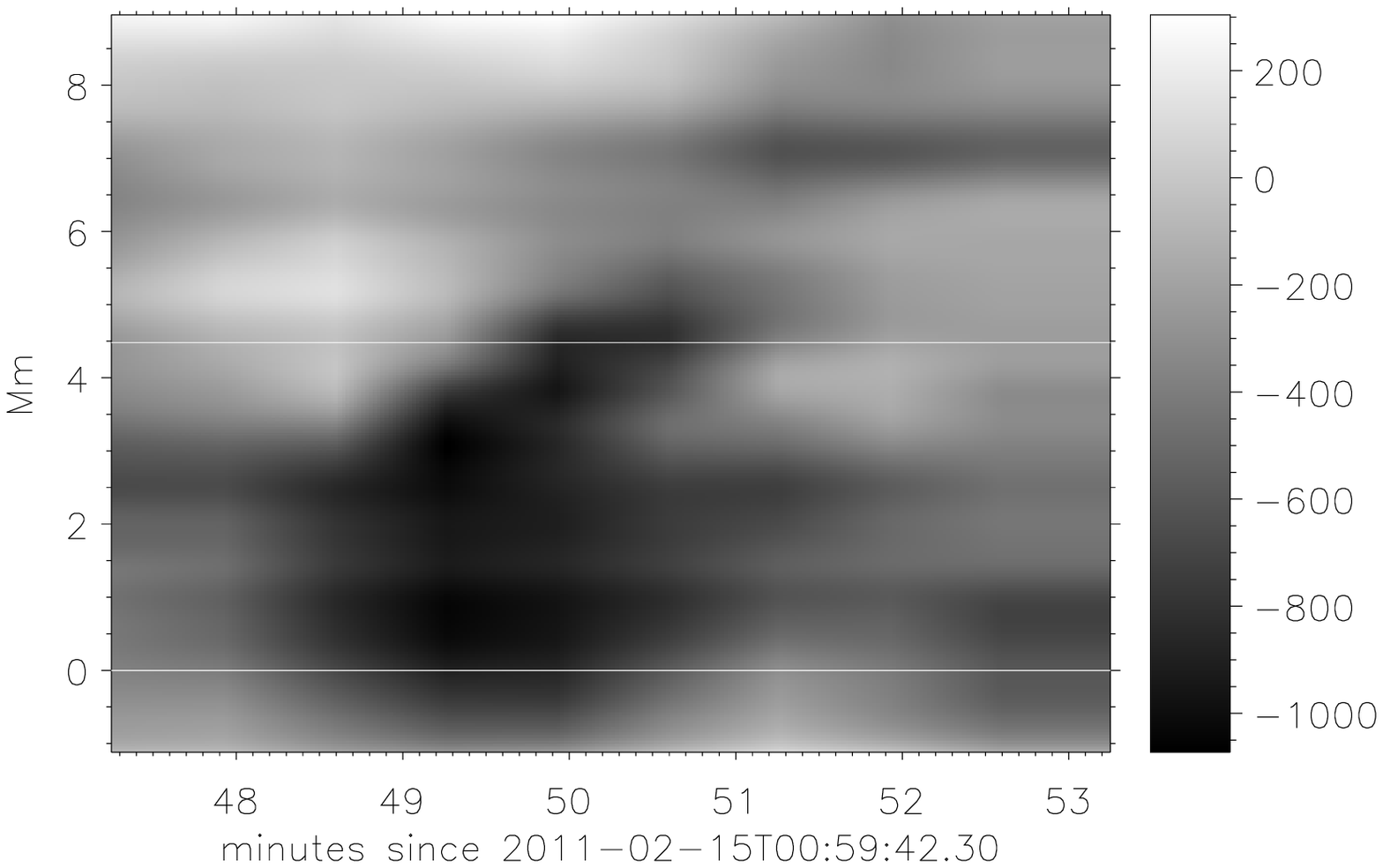} 
\end{tabular}
}
\caption{Source 1 (left) and 2 (right) velocity transients associated with the quake. Top row: HMI velocity data sequence around the source, the times in minutes since 01:00 UT are given in the title, location of time--distance sources is marked by orange star in right-most panels.
Middle row: 
HMI magnetogram (left) and velocity images with the lines along which the stack plots are made. The values of 0 and 1 correspond to horizontal lines (lower and upper correspondingly) in the middle row plots;  The red arcs represent angle of integration used for computing time--distance diagrams in Figure \ref{fig:td_s2}. Red stars mark locations of time--distance sources.
%	Middle row: stack plot from velocity data; 
Bottom row: stack plot from velocity  data. In all velocity plots, black and white correspond to downward and upward motions respectively. The black in magnetogram (middle row, left) corresponds to negative polarity, white to positive.}
\label{fig:transients_vd}
\end{figure*}

We use full disk SDO/HMI intensity, dopplergram, and line-of-sight magnetic field data at 45-second cadence to produce three hour-long datacubes, which are extracted by remapping and de-rotating the region of interest using Postel projection and the Snodgrass differential rotation rate. The spatial resolution is 0.04 degrees per pixel. The hard X-ray data  come from RHESSI \cite{lin02}, which observed the flare from the pre-cursor phase beginning at ~01:27\,UT until
02:30\,UT, covering the entire impulsive phase. We used the CLEAN algorithm to
produce images at between 20- and 40-second cadence covering the duration of the flare.  

We apply acoustic holography to calculate the egression power maps from observations. The holography method 
(\opencite{BL1999}; \opencite{Donea1999}; \opencite{BL2000a}; \citeauthor{LB2000}, \citeyear{LB2000,LB2004}) works by using Green's function [$G_+ (|\vect{r}-\vect{r}'|, t-t')$] which 
prescribes the acoustic wave propagation from a point source, to essentially ``backtrack'' the observed surface 
signal [$\psi(\vect{r}, t)$]. This allows us to reconstruct egression images showing the subsurface acoustic sources and sinks:
\be
H_+({\vect{r}, z, t})=\int \diff t' \int_{a<|\vect{r}-\vect{r}'|<b} \diff^2\vect{r}' G_+ (|\vect{r}-\vect{r}'|, t-t') \psi(\vect{r}', t'),
\label{equ:holo1}
\ee
where $a, b$ define the holographic pupil. The egression power is then defined as 
\be
P(z, \vect{r})=\int |H_+({\vect{r}, z, t})|^2 \diff t, 
\label{equ:holo2}
\ee
and can be viewed as a proxy for acoustic energy emitted from a location around $(z, \vect{r})$ over the integration period. In the above, $\vect{r}$ represents horizontal position, $z$ depth, and $t$ time.

In this work, the egression power is computed for each integral frequency from 3 to 10 mHz, by applying 2-mHz frequency bandwidth filters to the data (for acoustic energy estimates in Section \ref{sec:results} we use the same egression computation but with 1-mHz frequency bandwidth) and using Green functions built for surface monochromatic point source of corresponding frequency using geometrical optics approach \cite{Donea1999,donea2000,LB2000,MZZ2011,ZGMZ2011,ZH2011}. The pupil size is set from 10 to 40 Megameters. As flare acoustic signatures can be submerged by ambient noise for the relatively  long periods over which the egression power maps are integrated, we follow \inlinecite{DL2005} and use egression power ''snapshots'' to discriminate flare emission from the noise. Such a snapshot is simply a sample of the egression power within a time $\Delta t=\frac{1}{2 {\rm \  mHz}}=500 {\rm \ seconds}$. 

Time--distance diagrams (\opencite{kz1998}; \citeauthor{K2006} \citeyear{K2006,Kosovichev2007}; \opencite{Zharkova07}; \opencite{Kosovichev2011}; \opencite{ZH2011}) are computed by selecting a source location, rewriting the observed surface velocity signal [$v$] in polar coordinates relative to the source [$ v(r, \theta, t)$] and then using the azimuthal transformation 
\begin{equation}
V_m(r, t)=\int^\beta_{\alpha} v(r, \theta, t) {\mathrm e}^{-{\mathrm i} m \theta} \diff \theta, \label{equ:td}
\end{equation}
where integration is normally performed over the whole circle (i.e. $\alpha=0$, $\beta=2\pi$). In the case when $m=0,$  the integration limits  can be chosen to represent an arc. Acoustic wave-packets of a sufficiently strong amplitude relative to the noise are seen as a time--distance ridge that largely follows a theoretical time--distance curve.

The time--distance method is an observationally direct technique. If one is to think of circular ripples propagating from a source, at any moment of time, the procedure basically contracts ripples at a particular distance into a point on a time--distance diagram. Since only a handful of known sunquakes were accompanied by visible ripples, this obviously increases the chances of detection.
Nonetheless, the detection of the time--distance ridge can be affected by many factors such as noise in the data, filtering and image-processing techniques used. The acoustic egression method, on the other hand, provides a more quantitative measurement of the seismic source, but relies on a theoretical model of acoustic wave propagation from the source through solar interior to the surface. In a way, this method contracts the time--distance ridge into a single point measurement. 
Indeed, if we imagine a point source momentarily generating acoustic waves in an ideally quiet Sun, {\it e.g.} without convective motions,  then these waves will be observed as accelerating ripples on the surface, contracted to a ridge in a time--distance diagram computed at the source, and seen as a bright emission around the source time and location in an egression power map. 
 The holographic method is more sensitive, but vulnerable to noise, variations from the assumed model and is susceptible to an increased possibility of false detections, which is normally handled by analysing the statistical significance of the detected signal \cite{donea2000,DL2005,MZZ2011}.

\section{Results}
\label{sec:results}
%*******************************************************************************************************************
\begin{figure}
\centerline{
\begin{tabular}{c @ {\hspace{0pc}} c}
\includegraphics[width=7.cm]{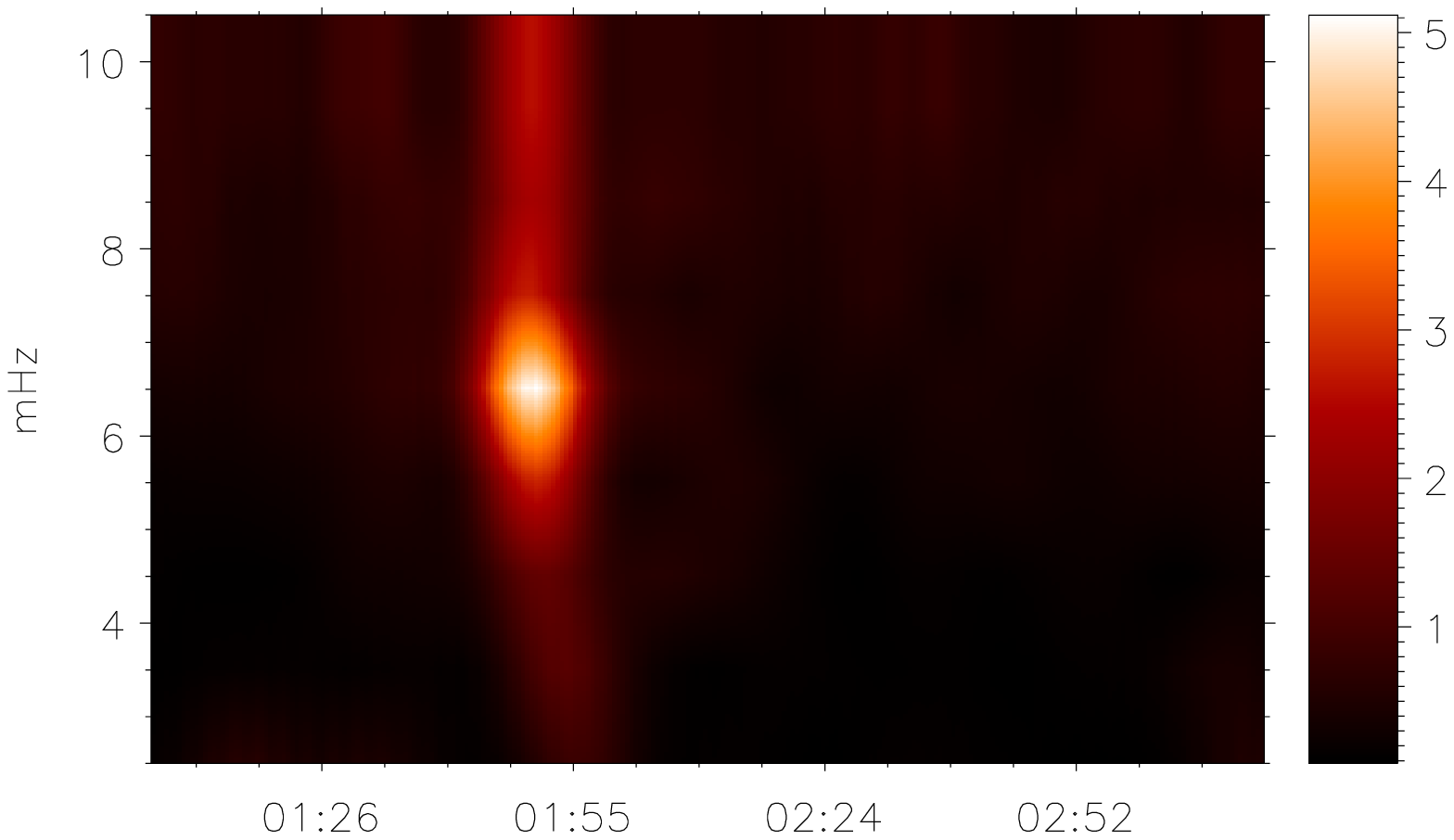} & 
\includegraphics[width=7.cm]{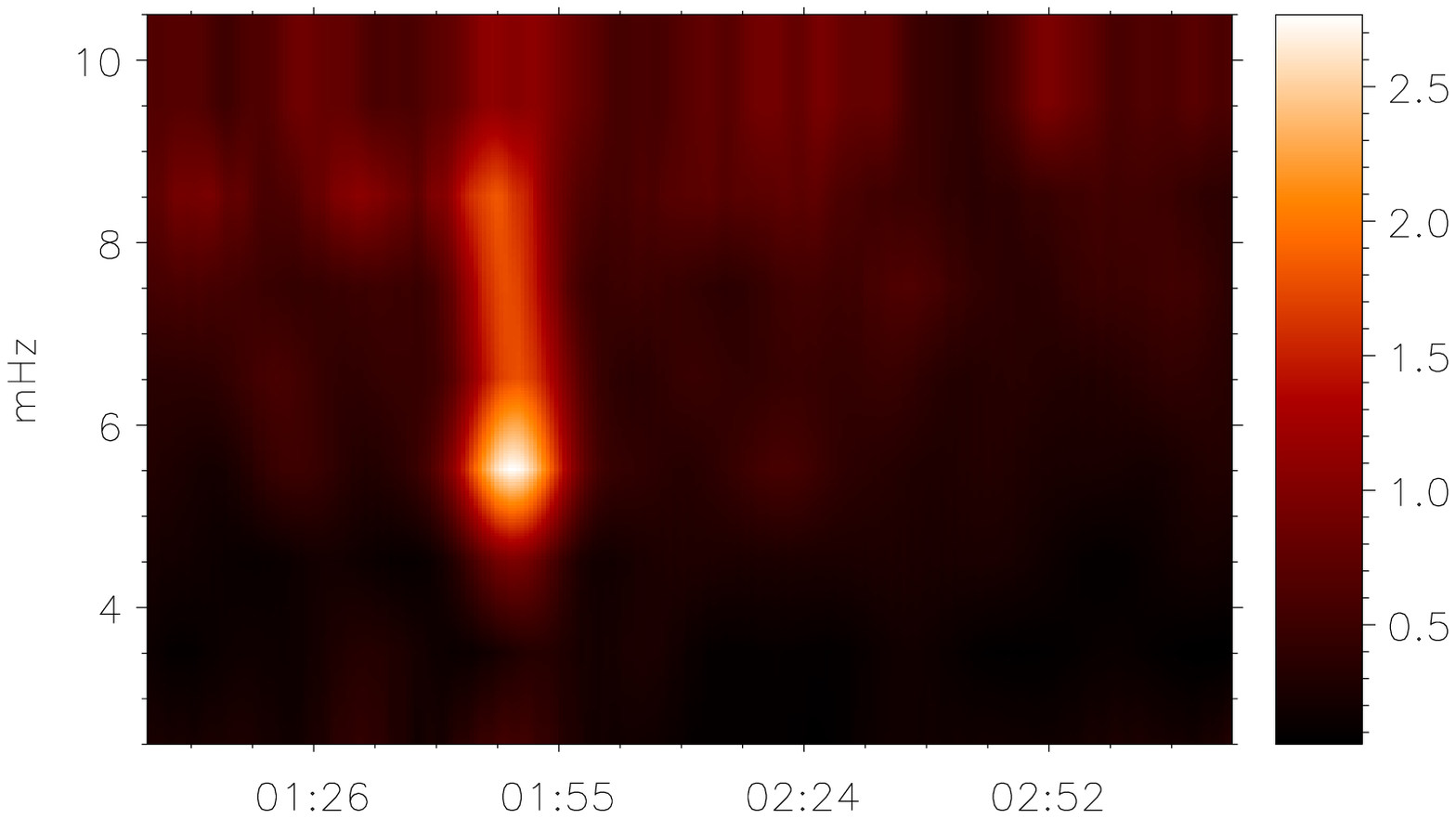}\\
\includegraphics[width=7.cm]{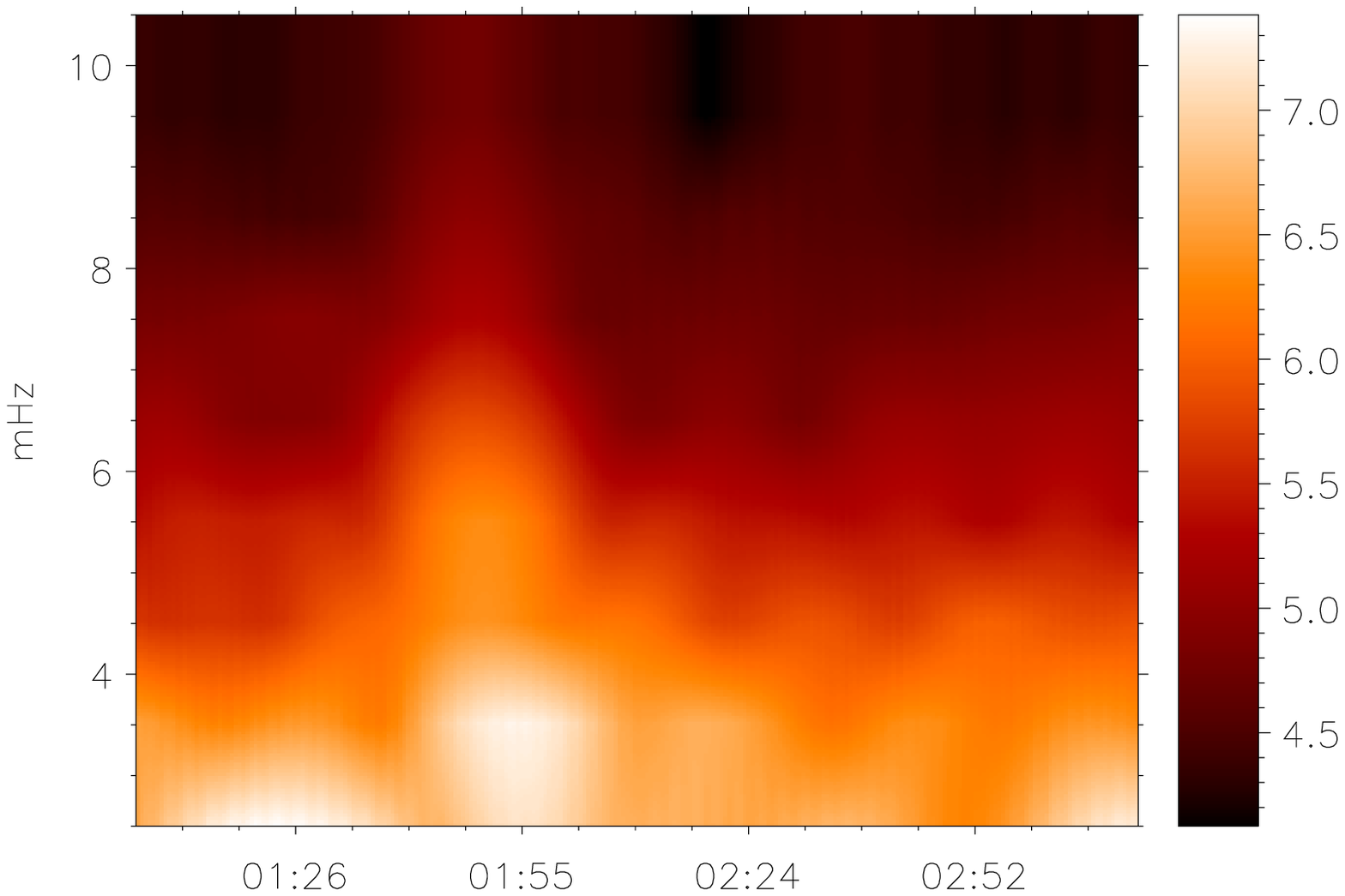} & %{solphys_rms_s1_logenergy_2.eps} &
\includegraphics[width=7.cm]{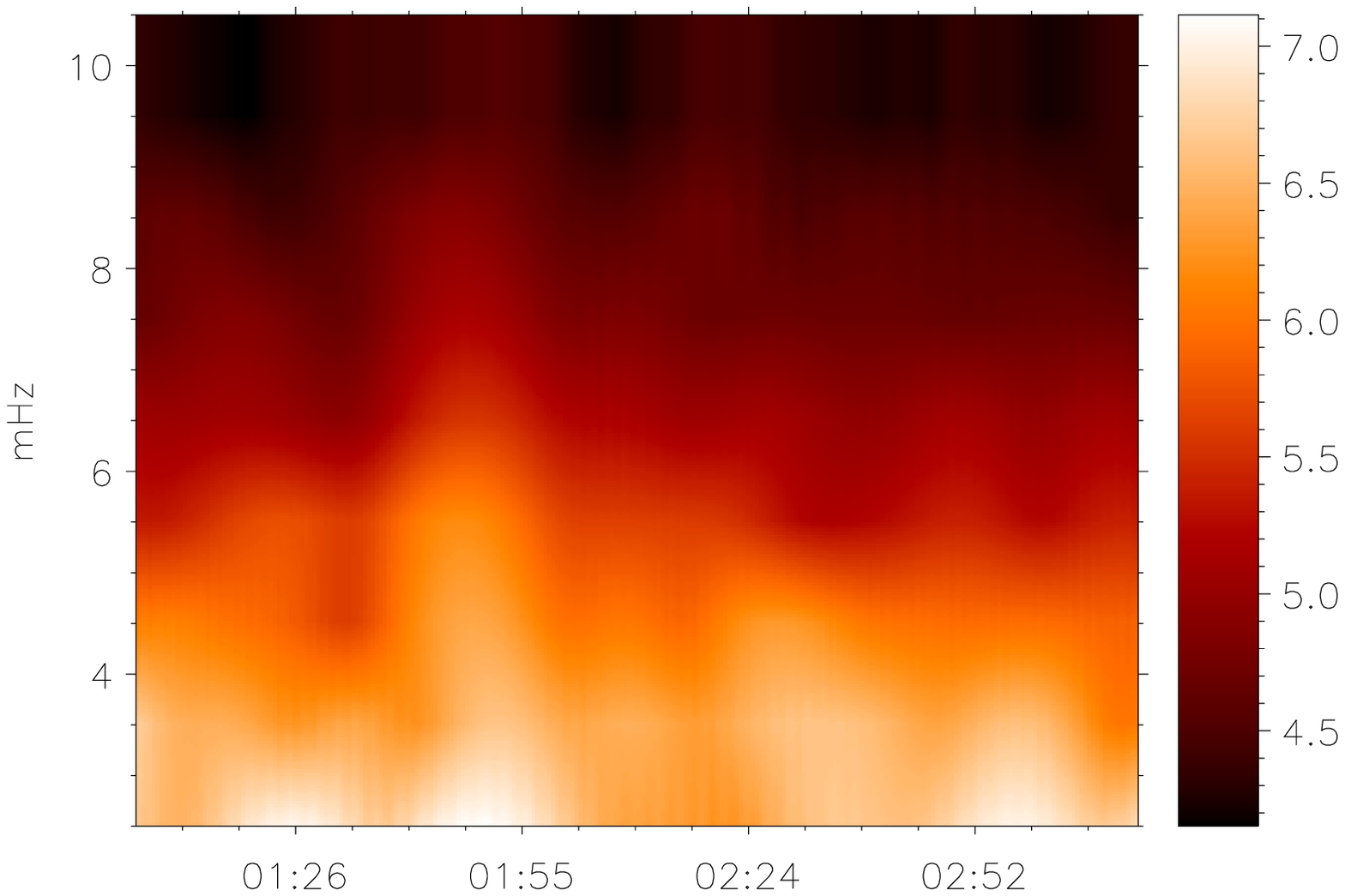}% {solphys_rms_s2_logenergy_2.eps}
\end{tabular}
}
\caption{Frequency response for Source 1 (left) and Source 2 (right). Egression power rms as function of time (along $x$-axis) and frequency (along $y$-axis) are plotted in units of quiet-Sun egression power (top row) and as $\log_{10}$ of acoustic energy flux [${\rm erg \,} {\rm cm^{-2}\,} {\rm s^{-1}}$]  (bottom row) for corresponding frequency band. 
} \label{fig:rms}
\end{figure}

%\subsection{Seismic sources}
Figure \ref{fig:srcs_boxes}(a) shows the egression power snapshot in the 6-mHz frequency band at around 01:51\,UT. The data is saturated at three times the quiet-Sun egression power for better contrast. The image shows the locations of the two seismic sources as bright regions. The larger source (on the left, East) corresponds to the one reported by \inlinecite{Kosovichev2011}.  We will call it Source 1. The smaller source on the right (West, Source 2) was detected by  \inlinecite{ZGMZ2011}.
Figures \ref{fig:srcs_boxes}(b)\,-\,(e) provide context for seismic observations by overplotting egression-power contours at 2.5 and 3 times the quiet-Sun egression units on HMI intensity, magnetogram and velocity data. 

In Figure \ref{fig:td_s2} we present time--distance diagrams computed from unfiltered HMI velocity difference data for both sources. Locations of the time--distance sources are marked by red stars in Figure \ref{fig:srcs_boxes}(b)\,-\,(e). Theoretical time--distance lines fitted to the data provide an estimate for the quake start times, suggesting that Source 2 is initiated at around 01:49:30 UT, about a minute earlier than Source 1.

Time--distance diagrams for both sources are computed in the direction of the surface velocity transient movement with integration in (\ref{equ:td}) performed over $90^\circ$ arc. These are shown in Figure \ref{fig:transients_vd} where we present the results of our analysis of the the horizontal motions of velocity transients at the quake locations. The upper plots are a time sequence of velocity data from small regions encompassing each seismic source, showing strong transient downflows and horizontal movement at the time--distance sources, which are marked by a star in the rightmost panel in each case. The middle row plots show red dashed arcs representing the $\theta$-integration range  used for computing time--distance diagrams in Figure \ref{fig:td_s2}  (see Equation (\ref{equ:td})). These are overplotted on HMI line-of-sight magnetogram and magnetogram-difference images. There we also define stack lines along which we present the velocity  variation in the bottom row. The time is along the $x$-axis, and horizontal distance is along the $y$-axis.
In each case, the point marked as zero, corresponds to the lower horizontal line (at $y=0$) in the bottom row, with 1 corresponding to the upper horizontal line.

For Source 1 (left column in Figure \ref{fig:transients_vd}) we can see a strong downflow  of about $500-700 {\rm \ m \ s}^{-1}$ starting around 01:50\,UT followed by the upflow, travelling around 2.5 Mm along the line in around two\,-\,three minutes. This gives a speed estimate between $13-20 {\rm \ km \ s}^{-1},$ which is around the value reported by \inlinecite{Kosovichev2011}. Source 2 (right column in the Figure  \ref{fig:transients_vd}) shows a persistent strong downflow of around $1 {\rm \ km \ s}^{-1}$ located near the zero point from around 01:48\,UT to 01:52\,UT as well downward transient movement along the line. The horizontal movement starts about 01:48:30\,UT and moves with the estimated speed $14-22 {\rm \ km \ s}^{-1}.$

\begin{figure}
\centerline{
\includegraphics[width=\textwidth]{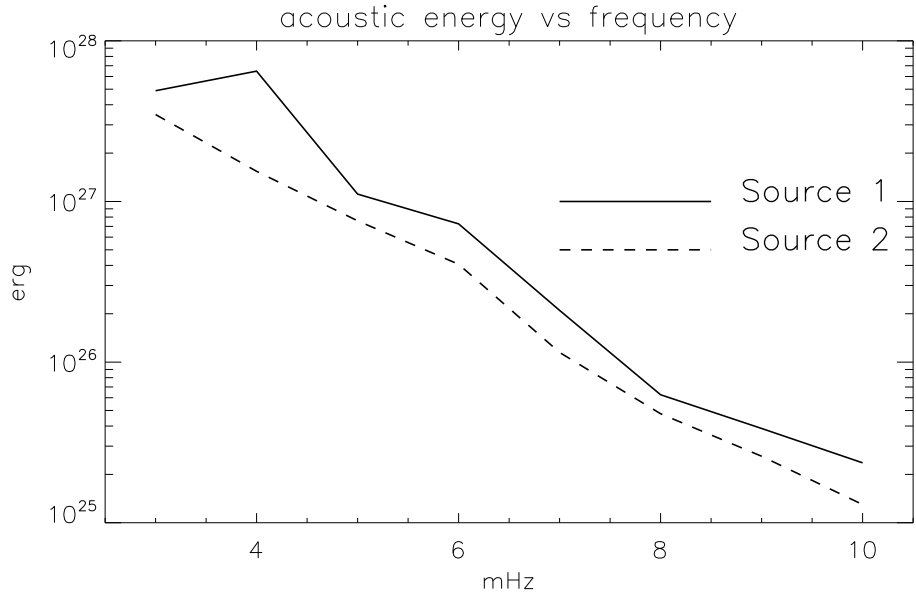}%{solphys_energy_vs_frequency}
}
\caption{Acoustic energy distribution {\it versus} frequency for the two sources. 
%Thin lines correspond to estimates of acoustic energy emitted over the same time period and over the same area as those of the two sources. 
Solid line corresponds to Source 1, dashed to Source 2.}
\label{fig:energy_vs_freq}
\end{figure}

To measure the acoustic energy for the seismic sources we have recomputed the acoustic egression as described above but applying 1-mHz bandwidth filtering, instead of 2 mHz, so that the measurements at each frequency band do not overlap. We used a 6-mHz egression-power snapshot to define kernels for spatial integration by thresholding at the quiet-Sun intensity and selecting congruent areas corresponding to sources one and two. Then, at each frequency, the egression power scaled by the sound-speed and density values taken from Model C \cite{CDModel} at around 250 km above the photosphere was integrated over such kernels and a 30-minute period around the peak of HXR emission. The results are presented in Figure \ref{fig:energy_vs_freq}.  
From this we estimate that the total amount of acoustic energy released by Source 1 is around   $1.18\times 10^{28} {\rm \ ergs}$ and Source 2 is around $6.08 \times 10^{27} {\rm \ ergs}$.     

Figure \ref{fig:rms} shows the rms variation of egression power spatially integrated around the two seismic sources as a function of frequency and time. In the top row of Figure \ref{fig:rms}, the egression power is expressed in units of quiet-Sun egression power computed for the corresponding frequency, with the log of the integrated egression in   ${\rm erg\, cm^{-2}\, s^{-1}}$ presented in the bottom row. 
%From the Figure 
The relative strength of the eastern source (Source 1, left column) is apparent, with a significant increase in relative egression power seen in all frequencies. The western source (Source 2, right column) has a clear signature around 4\,-\,7 mHz range, with little apparent power increase around 3\,mHz or frequencies above 9\,mHz.  The egression method does not provide an accurate determination of the quake onset time
due to the frequency-bandwidth filtering of the data and the induced timing uncertainty of 500 or 1000 seconds for 1-mHz and 2-mHz filters respectively \cite{Donea1999,DL2005}. Nonetheless, it is interesting to note from Figure \ref{fig:rms} the relative frequency variation of egression peak times.

Figure \ref{fig:td_src_vars} shows the photospheric changes in HMI line-of-sight velocity, magnetic field, and continuum intensity at the time--distance sources (indicated in Figure \ref{fig:srcs_boxes}) over two hours  (top panel) and 40 minutes (bottom panel) around the flare. 
                                                                                                                                                                                                                                                                                                                                     
%**********************************************************************************************************

\section{Discussion}
\label{sec:discussion}

%*******************************************************************************************************************

%{\bf Strictly speaking  there are no acoustic waves in sunspots by definition, only magneto-acoustic waves.} 
When magnetic field is present, as is the case for sunspots, the acoustic waves become magneto-acoustic.
Numerical and theoretical modelling of acoustic-wave packet propagation through a magnetised plasma \cite[for example]{C2000,Schunker2006,SZ09,Moradi2010,Felipe2010} shows that acoustic waves when passing through magnetic-flux tube transform into different types of magneto-acoustic waves, which we then expect to find in sunspots. 
Mode conversion between various magneto-acoustic waves takes place around the region where the ambient sound speed [$c$] equals the Alfven velocity [$a$] which is situated near the surface in sunspots. 
For instance, the slow-mode waves (classified as being slow below the $a=c$ transition region) have properties fundamentally different from those expected from non-magnetic acoustic waves. These waves propagate along magnetic-field lines and therefore have very different wavefronts and speeds. 
However, as shown by \inlinecite{SZ09}, the fast-mode magneto-acoustic waves are similar to acoustic waves. While the presence of a magnetic field and associated changes in ambient conditions clearly affect their paths and speed, these changes are relatively small.  When travelling outside the magnetic tube, these waves become acoustic. 

While the physical mechanism behind the excitation of the waves remains undetermined, we can hypothesise that the observed flare-generated acoustic wavefronts are likely connected to such magneto--acoustic waves. Indeed, as quakes happen in magnetised plasma, the generated waves will be magneto-acoustic. However, the acoustic nature of the ripples and ridges in the time--distance diagrams is very clear:  detected ridges closely follow the theoretical travel time (see 
\citeauthor{K2006} (\citeyear{K2006,Kosovichev2007,Kosovichev2011}); \inlinecite{Martinez2008a}, {\it e.g.}). Nonetheless, the subsurface conditions will affect the time and speed of the propagation of such waves, introducing an anisotropy of the wavefront such as that reported by \inlinecite{Kosovichev2011} for Source 1. As the egression measurements are based on a quiet-Sun model of acoustic-wave propagation, they are certainly affected by such an anisotropy. 
On the other hand, the travel time perturbations of acoustic waves passing through sunspots measured by time--distance helioseismology \cite{Duvall1993,Duvall1997,Kosovichev2000,ZNT2007} indicate that such changes are expected to be small and of the order of one minute.

\begin{figure*}
\centerline{
\begin{tabular}{c}
\includegraphics[width=1.1\textwidth]{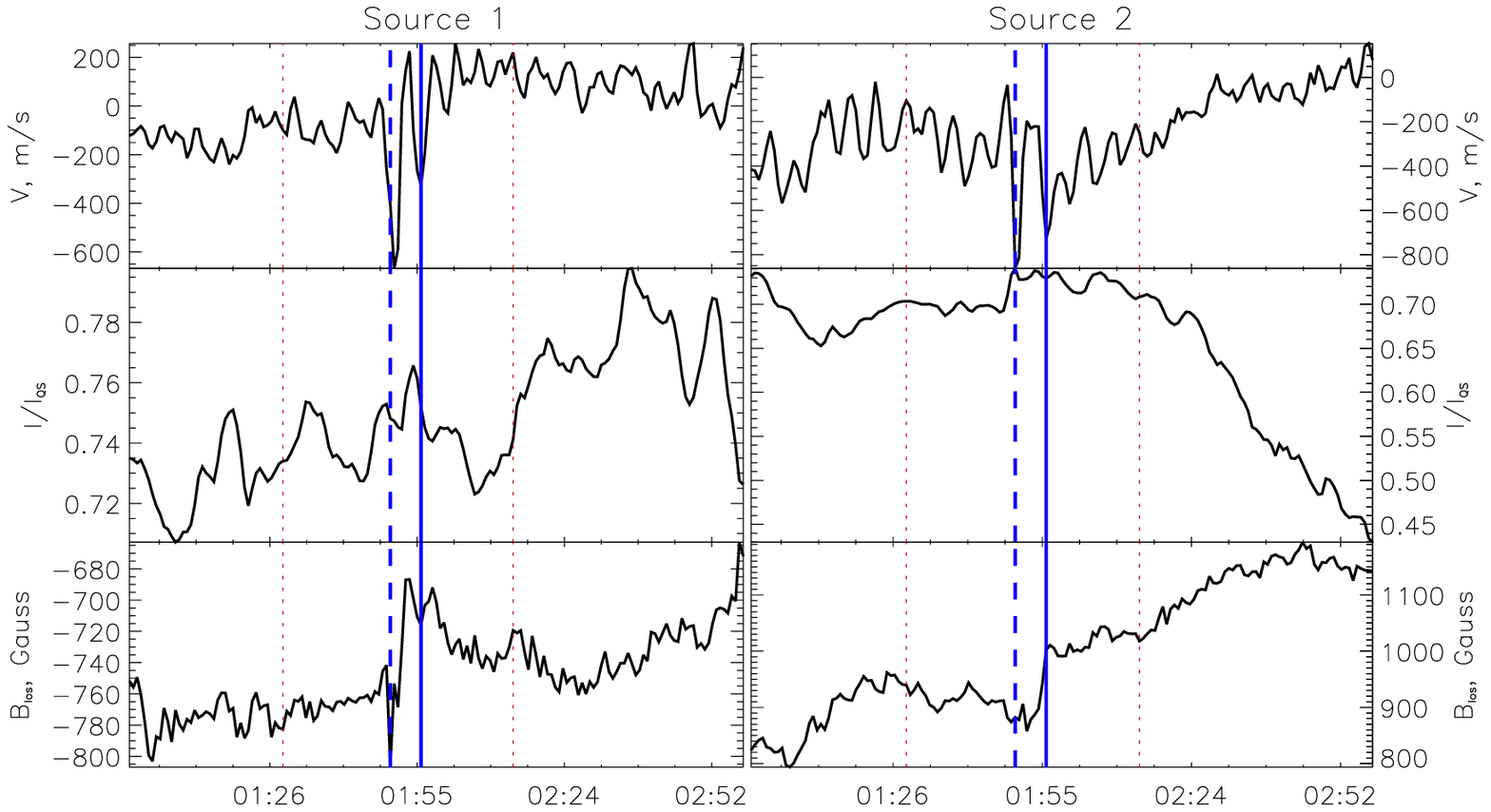} \\
\includegraphics[width=1.1\textwidth]{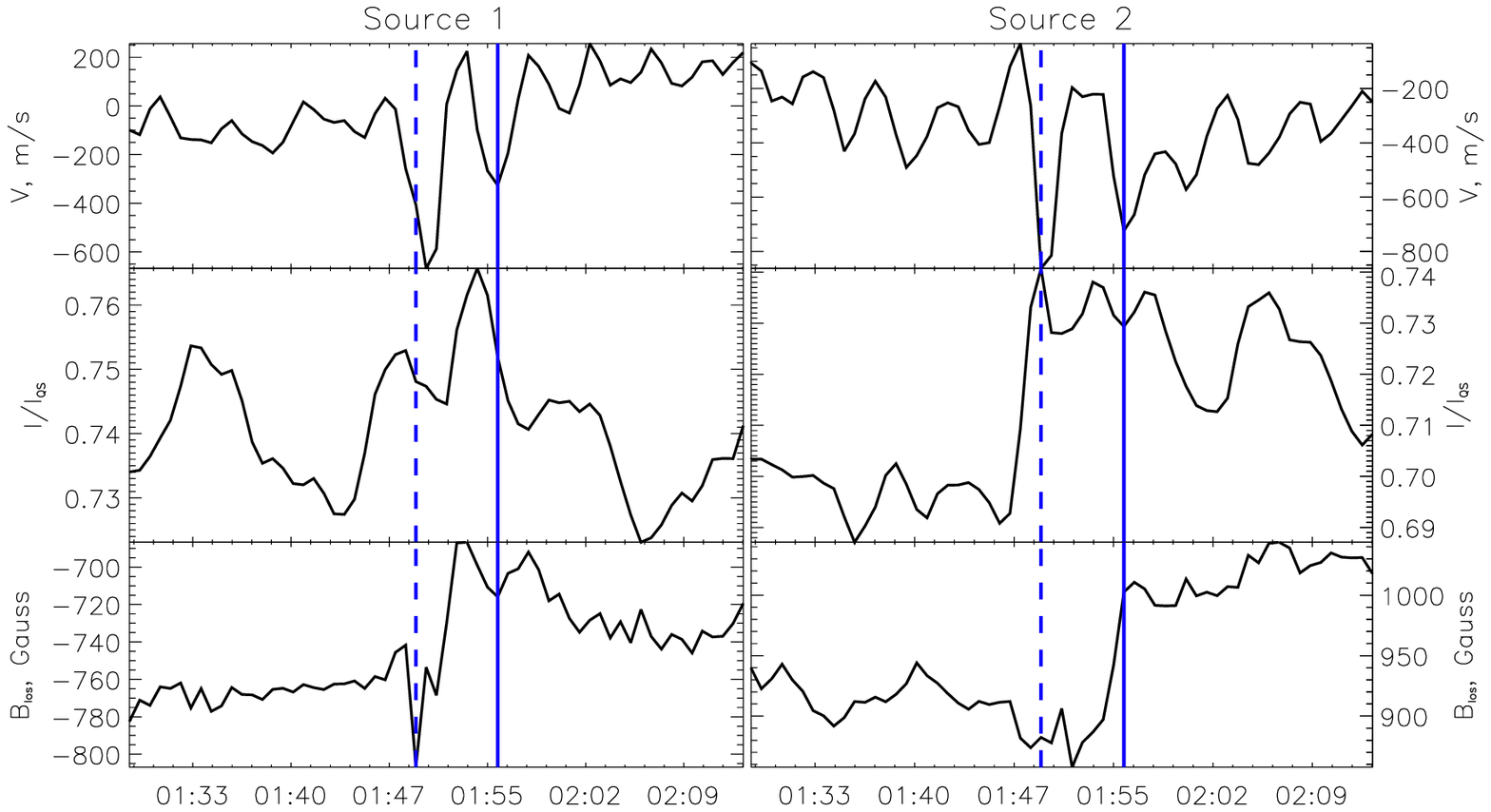} 
\end{tabular}
}
\caption{Variations of velocity, intensity, and magnetic field at time--distance sources 1 (left) and 2 (right); Top plots show two-hour time series around the time of the flare, the bottom plots are the same but on a 45 minute scale. The red dotted line in the top plots indicate the edges of the bottom plot for reference. The vertical dashed and solid blue lines correspond to 01:49:57 and 01:55:57\,UT respectively. Please see text for more details. 
}
\label{fig:td_src_vars}
\end{figure*}

The stacked nature of egression kernels presented in Figure \ref{fig:srcs_boxes}(a), is very similar to the measurements obtained for other quakes \cite{Donea1999,DL2005,donea11}. In these works the authors suggest that this could be the result of interference caused by the rapid motion of the source, roughly in the direction along which the kernels are stacked. For example, \inlinecite{DL2005} found that for 28 and 29 October 2003 quakes the motion of the HXR sources was indeed aligned accordingly with egression power stacks. This was also  confirmed for the 23 July 2002 flare by \inlinecite{Kosovichev2007} using HXR and Doppler data.

For this flare, 15 February 2011, \inlinecite{Kosovichev2011}  used time--distance diagram analysis to detect a  supersonic (adiabatic sound speed at the photospheric level is around $7-8 {\rm \ km \ s}^{-1}$ in the quiet Sun) movement of the location of Source 1 of around $15-17 {\rm \ km \ s}^{-1}$. In fact, apparent horizontal motions of downward transients are observed in the HMI velocity-data sequence at the location of both seismic sources before and around the time of the quakes (see Figure \ref{fig:transients_vd} for instance). In order to investigate this, we have produced stack plots (bottom row of Figure \ref{fig:transients_vd}), from which the horizontal speed in the direction along the integration line is estimated  (Section \ref{sec:results}). While the sound speed in the penumbral magnetised plasma is  likely to be different from the quiet Sun, given the estimates ($13-20$ and $14-22  {\rm \ km \ s}^{-1}$ for sources 1 and 2 respectively)  it is reasonable to conclude that in both cases the motion appears to be supersonic. We note that supersonic horizontal movements in the photosphere have also been reported in association with several other flare induced sunquakes (see for example \citeauthor{DL2005}, \citeyear{DL2005}; \citeauthor{Kosovichev2007}, \citeyear{Kosovichev2007}).  

Considerable anisotropy in the acoustic amplitude of the ripples from the vantage of the sources has been observed for most quakes \cite{K2006,Moradi2007,donea11}. In fact, \inlinecite{donea11} suggests that the maximum amplitude of the ripples emanating from a moving source is generally along the axis of the source, displaced from the source location in the direction of the motion. This is confirmed for both of our sources by our directional time--distance diagrams (see Figures \ref{fig:td_s2} and  \ref{fig:transients_vd}, where after considering various arcs, the results where the time--distance ridge appears the strongest are presented).
%After considering various arcs, the results where the time--distance ridge appears the strongest are presented in Figure \ref{fig:td_s2}.

As mentioned above, theoretical time--distance curves fitted to the ridge data in the time--distance diagrams suggest a marginally earlier start for the western quake. As this difference is within a margin of error, we consider velocity plots from both time--distance sources  shown in Figure \ref{fig:td_src_vars}. These indicate strong downflows around the times of the quake. It is also clear that the Source 2 downflow happens about 45 seconds earlier than Source 1. This confirms the timing difference between the two sources.

%\subsection{Photospheric transients and changes associated with the flare}
The plots for the magnetic-field and intensity variation over two hours for Source 2 in Figure \ref{fig:td_src_vars} show a long-term continuous increase in magnetic-field flux density coupled with the intensity decrease, which indicates ongoing emergence of the flux at the location. Around the time of the flare, however, we see a gradual and apparently permanent magnetic-field increase of around $130-160$ Gauss, taking place from around 01:50 to 01:56\,UT, coinciding with the peak of the flare X-ray emission. This is preceded by a $7\%$ increase in intensity taking place approximately from 01:47 to 01:50\,UT. After its peak, the intensity decreases slowly to a pre-flare level, suggesting that a physical process not related to sudden heating, but perhaps associated with the magnetic-field change, is taking place. There is also present a transient magnetic-field variation of about $50$ Gauss at around 01:51\,--\,01:52 that is likely to be associated with the abrupt changes in the observed line profile due to precipitation of energetic particles. An abrupt increase in the line-of-sight magnetic field could indicate that flux-rope field lines have become more vertical, which would be in agreement with the eruption model suggested by \inlinecite{ZGMZ2011}.  Source 1 plots, on the other hand show a transient increase in the magnetic-field strength of about 50\,--\,60 Gauss around 01:50 followed by a rapid decrease with the field settling back to pre-flare levels about half an hour later. There is also a very small increase in continuum intensity increase of around $2.7\%$.

Similar plots of photospheric variations where the data were averaged over the egression sources have also been presented in Figure 4 of \inlinecite{ZGMZ2011}. We note the difference between the two figures in the magnetic-field and intensity responses. We find that these parameters vary significantly from pixel to pixel in egression kernels, suggesting that it is very important to pinpoint the exact location of the quakes. 
These can be determined more precisely using time--distance diagrams, as pixels with respect to which the diagrams showing a clear ridge are computed form a connected set for each source.
%{\bf While time--distance diagrams allow us to determine the location the quakes more precisely, we find that both sources can be defined as being a connected set of pixels where the time--distance diagram computed with respect to each pixel shows a clear ridge.} 
However, we see different magnetic-field behaviour at the seismic location from pixel to pixel: for example, a few pixels show signs of a permanent magnetic change, many show an oscillatory kind of behaviour, and even more show a transient-like response. Indeed, the results of our preliminary analysis suggest that abrupt and permanent magnetic changes are more prevalent in the ribbons, although the picture is still quite complex, perhaps due to the continuing flux emergence in the region. 
Further analysis making use of known methods \cite{Sudol2005,Zharkova2005} and now available HMI vector magnetic data is needed to understand the variability of magnetic-field changes in seismic sources and this flare in general.

The acoustic-energy estimates obtained here are somewhat higher than obtained for the same flare by \inlinecite{ABMLHC2012}. This is most likely due to several factors such as use of Model C as opposed to VAL, different pupil dimensions and different integration kernels (we use larger areas around 37 and 33 ${\rm Mm^{2}}$ for Sources 1 and 2 correspondingly, as opposed to 12 ${\rm Mm^{2}}$  (\inlinecite{ABMLHC2012}, C.~Lindsey, private communication, 2012). The estimate for Source 1 puts it amongst the most powerful sunquakes associated with X-class flares \cite{Moradi2007,Besliu2005}. We note, however, that \inlinecite{Besilu2006Mclass} reports an even stronger seismic event for the M7.8 flare on 2 December 2005. 

\section{Conclusions}

We have presented here a time--distance diagram in addition to the one found by \inlinecite{Kosovichev2011} showing a clear ridge for the second seismic source associated with the 15 February 2011 X-class flare. Using time--distance analysis and HMI line-of-sight velocity observations we deduce that the quakes are excited at around 01:50\,UT, with the eastern source onset preceding the western one by about $45-60$ seconds. We have also detected apparent horizontal motions of the downward velocity transients at the time and location of both quakes. The speed of such motions is larger than the ambient sound speed. The direction of such motions is  aligned with the stacked egression kernels and amplitude anisotropy of the generated wavefront, indicating that a moving source is the likely scenario for both quakes. 
We estimate the acoustic energy released by both quakes to be around $1.18\times 10^{28} {\rm \ ergs}$ for Source 1 and $6.08 \times 10^{27} {\rm \ ergs}$ for Source 2. For Source 1 this is about an order of magnitude higher than the Lorentz-force energy estimate for a generic flare provided by \inlinecite{Hudson08}. This is in line with findings by \inlinecite{ABMLHC2012}, where more accurate evaluation of Lorentz-force energy has been produced. However, given a number of simplifications used in obtaining the Lorentz-force estimate, such as the use of line-of-sight magnetic field only, the assumption of a single area where changes occur and relatively low magnetic-field strength, in our view it would be premature to discard the Lorentz force as a possible mechanism for quake excitation with further analysis based on \cite{Fisher2011,Fisher2012} making use of fill vector magnetic-field data necessary.

Further analysis is required in order to understand the physical nature of both detected quakes and their link with the flux-rope eruption that was associated with the X-class flare. In particular, the HMI vector magnetogram data should shed the light on full magnetic changes, and numerical extrapolations of the three-dimensional magnetic field from photosphere through atmosphere and corona will give us a clearer picture of the magnetic-field restructuring and energy release associated with this event.

\acknowledgements
The authors thank C. Lindsey for his help in obtaining acoustic-energy estimates. The authors also thank the anonymous referee for their insightful comments. We acknowledge the Leverhulme Trust for funding the ''Probing the Sun: inside and out'' project upon which this research is based.

%*******************************************************************************************************************

%\bibliographystyle{spr-mp-sola}
%\bibliography{analysis20110215}{}
%\input{solphys_rhodes_revision_2refs_1.tex}

\end{article}
\end{document}